\documentclass[11pt]{article}

\usepackage{geometry}
\usepackage[round]{natbib}
\usepackage{enumitem}
\usepackage{cite}
\usepackage{graphicx}
\usepackage{subfigure}
\usepackage{hyperref}
\usepackage{amssymb}
\usepackage{amsmath}
\allowdisplaybreaks[1]

\usepackage{amsthm}

\newtheorem{theorem}{Theorem}
\newtheorem{lemma}[theorem]{Lemma}

\makeatletter
\def\@endtheorem{\endtrivlist}
\makeatother

\newcounter{rrule}
\newenvironment{rrule}{\refstepcounter{rrule}\par\smallskip\noindent
\textbf{Rule \arabic{rrule}.}\quad}{}

\newcommand{\edge}[2]{(#1,#2)}
\newcommand{\und}[1]{\mathrm{Und}(#1)}
\newcommand{\C}{\mathcal{C}}
\newcommand{\D}{\mathcal{D}}
\newcommand{\PP}{\mathcal{P}}
\newcommand{\Vd}{V_{\text{del}}}

\begin{document}

\title{A smaller kernel for Ptolemaic Vertex Deletion}
\author{Dekel Tsur%
\thanks{Faculty of Computer and Information Science, Ben-Gurion University of the Negev.}}
\date{}
\maketitle

\begin{abstract}
In the \textsc{Ptolemaic Vertex Deletion} problem, the input consists of an undirected graph $G$
and an integer $k$.
The goal is to decide whether there is a set of at most $k$ vertices
whose removal from $G$ results in a Ptolemaic graph.
In this paper, we give a kernel for \textsc{Ptolemaic Vertex Deletion} with
$O(k^5)$ vertices.
Our result improves upon the $O(k^6)$ kernel of Agrawal et al.~[IPEC 2021].
\end{abstract}


\section{Introduction}

Let $\Pi$ be a class of graphs.
In the \textsc{$\Pi$ Vertex Deletion} problem, the input consists of an undirected graph $G$
and an integer $k$.
The goal is to decide whether there is a set of at most $k$ vertices
whose removal from $G$ results in a graph belonging to $\Pi$.
Many vertex deletion problems have been studied in the computer science literature,
for example, \textsc{Feedback Vertex Set} (where $\Pi$ is the class of acyclic graphs),
\textsc{Vertex Cover} (where $\Pi$ is the class of edgeless graphs), and
\textsc{Cluster Vertex Deletion} (where $\Pi$ is the class of graphs in which every connected component
is a clique).

In this work, we study the class of Ptolemaic graphs.
A graph $G$ is a \emph{Ptolemaic graph} if for every four vertices $u,v,w,x$
in the same connected component of $G$, the following inequality holds:
\[d(u,v)d(w,x)+d(u,x)d(v,w) \geq d(u,w)d(v,x), \]
where $d(y,z)$ denotes the distance between vertices $y$ and $z$.
\citet{agrawal2021polynomial} gave a kernel for \textsc{Ptolemaic Vertex Deletion}
with $O(k^6)$ vertices.

In this paper, we give a kernel for \textsc{Ptolemaic Vertex Deletion} with
$O(k^5)$ vertices.
Our kernel is based on the kernel of \citet{agrawal2021polynomial}
and uses ideas from the kernel of \citet{Tsur_3leaf} for \textsc{3-Leaf Power Vertex Deletion}.

The kernel of Agrawal et al.\ works by constructing a set $A$ such that
$G-A$ is a Ptolemaic graph and $|A| = O(k)$, and then
considering the maximal cliques in the graph $G-A-R$, where $R$ is a set of $O(k^5)$ vertices
constructed by the algorithm.
Their kernel reduces the number of maximal cliques in $G-A-R$ by considering the inter-clique graph of $G-A-R$
(see Section~\ref{sec:preliminaries}).
Using reduction rules, the number of vertices in the inter-clique graph is reduced to $O(k^5)$,
providing a bound of $O(k^5)$ on the number of maximal cliques in $G-A-R$.
The kernel uses additional reduction rules to bound the sizes of the maximal cliques.
After applying these rules, there are $O(k^3)$ maximal cliques with $O(k^3)$ vertices each and
$O(k^5)$ maximal cliques with $O(k)$ vertices each,
excluding $O(k^5)$ \emph{special vertices}.
Thus, the number of vertices in $G$ is $O(k^3 \cdot k^3 + k^5 \cdot k + k^5) = O(k^6)$.

In contrast, our kernel considers a partition of the vertices of $G-A-R$ into clique-modules,
based on the inter-clique graph of $G-A-R$.
We show that the number of vertices of the inter-clique graph can be reduced to $O(k^4)$,
and therefore the number of clique-modules in the partition is $O(k^4)$.
We then use a reduction rule that reduces the sizes of the clique-modules.
After applying this rule, there are $O(k^3)$ clique-modules with $O(k^2)$ vertices each,
and $O(k^4)$ clique-modules with $O(k)$ vertices each,
excluding $O(k^5)$ special vertices.
Therefore, the number of vertices in $G$ is $O(k^3 \cdot k^2 + k^4 \cdot k + k^5) = O(k^5)$.

\paragraph{Related work}

The class of Ptolemaic graphs is the intersection of chordal graphs and distance-hereditary graphs.
For \textsc{Chordal Vertex Deletion},
\citet{marx2010chordal} showed that the problem is fixed-parameter tractable.
\citet{cao2016chordal} later gave a faster parameterized algorithm.
Regarding kernelization,
\citet{jansen2018approximation} gave a kernel with $O(k^{161}\log^{58} k)$ vertices,
and \citet{agrawal2018feedback} gave a kernel with $O(k^{12}\log^{10} k)$ vertices.
For \textsc{Distance-hereditary Vertex Deletion},
\citet{eiben2018single} gave an $O^*(37^k)$-time algorithm, and
\citet{kim2021polynomial} gave a kernel with $O(k^{30}\log^5 k)$ vertices.

Another related class of graphs is the 3-leaf power graphs,
which is a subset of the class of Ptolemaic graphs.
For \textsc{3-Leaf Power Vertex Deletion},
\citet{dom2006error} gave an $O^*(6^k)$-time algorithm.
Regarding kernelization,
\citet{ahn2023polynomial} gave a kernel with $O(k^{14})$ vertices,
and \citet{Tsur_3leaf} later gave an improved kernel with $O(k^6)$ vertices.

\section{Preliminaries}\label{sec:preliminaries}

For a set of sets $\mathcal{S}$, $\bigcup \mathcal{S} = \bigcup_{S \in \mathcal{S}} S$.
A \emph{partition} of a set $S$ is a collection of pairwise disjoint sets $S_1,\ldots,S_d$ such
that $S_1 \cup \cdots \cup S_d = S$ (note that some of the sets $S_i$ can be empty).

Let $G$ be an undirected graph.
For a vertex $v \in V(G)$, $N_G(v)$ is the set of its neighbors and $N_G[v] = N_G(v) \cup \{v\}$.
We may omit the subscript $G$ if the context is clear.
For a set of vertices $S \subseteq V(G)$, $N_G(S) = (\bigcup_{v \in S} N_G(v)) \setminus S$.
For a set of vertices $S \subseteq V(G)$, $G[S]$ is the subgraph of $G$ induced by $S$ and
$G-S = G[V(G) \setminus S]$.
For a vertex $v$, we write $G-v$ instead of $G-\{v\}$.
For a set of edges $F \subseteq E(G)$, $G-F$ is the graph $(V(G), E(G) \setminus F)$.
For two disjoint sets of vertices $A$ and $B$, we say that
\emph{$A$ and $B$ are adjacent} if there is at least one edge with one endpoint in $A$
and one endpoint in $B$.
A node of an acyclic undirected is called a \emph{leaf} if its degree is at most one.

Let $S$ be a set of vertices in a graph $G$.
$S$ is a \emph{clique} if all pairs of distinct vertices from $S$ are adjacent.
$S$ is an \emph{independent set} if no two distinct vertices from $S$ are adjacent.
$S$ is a \emph{module} if $N(u) \setminus S = N(v) \setminus S$ for all $u,v \in S$.

For a directed graph $G$, $\und{G}$ denotes the underlying undirected graph of $G$.
That is, $V(\und{G}) = V(G)$ and $\edge{u}{v} \in E(\und{G})$ if and only if
$\edge{u}{v} \in E(G)$ or $\edge{v}{u} \in E(G)$.

\subsection{Ptolemaic graphs}

Let $S$ be a set of vertices in a graph $G$.
$S$ is a \emph{deletion set} of $G$ if $G-S$ is a Ptolemaic graph.
$S$ is a \emph{solution} of $(G,k)$ if $S$ is a deletion set of $G$ and $|S| \leq k$.

\begin{figure}
\centering
\includegraphics{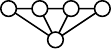}
\caption{A gem.\label{fig:gem}}
\end{figure}

We say that a graph $G$ \emph{contains} a graph $H$ if $H$ is an induced subgraph of $G$.
A graph $G$ is a \emph{gem} if it is isomorphic to the graph in Figure~\ref{fig:gem}.
The \emph{center} of a gem $G$ is the unique vertex of $G$ with degree~4.
A graph $G$ is a \emph{hole} if $G$ is a chordless cycle of length at least four.
A graph $G$ is an \emph{obstruction} if $G$ is a gem or a hole.

\begin{theorem}[\citet{howorka1981characterization}]\label{thm:ptolemaic-1}
A graph $G$ is Ptolemaic if and only if it does not contain an obstruction.
\end{theorem}

\begin{theorem}[\citet{ahn2022towards}]\label{thm:approximation}
There is a polynomial time approximation algorithm for \textsc{Ptolemaic Vertex Deletion}
with an approximation ratio of 68.
\end{theorem}

\begin{lemma}[\citet{agrawal2021polynomial}]\label{lem:cycle}
Let $G$ be a graph with $|V(G)| \geq 5$ and let $K$ be a Hamiltonian cycle in $G$.
If there is a vertex $v \in V(G)$ such that every edge in $E(G) \setminus E(K)$
is incident to $v$, then $G$ contains an obstruction that includes $v$.
\end{lemma}

\begin{lemma}\label{lem:cycle-2}
Let $G$ be a graph with $|V(G)| \geq 5$ and let $K$ be a Hamiltonian cycle in $G$.
If there are distinct vertices $v,w \in V(G)$ such that $\edge{v}{w} \notin E(G)$
and every edge in $E(G) \setminus E(K)$ is incident to $v$, then
$G$ contains a hole $H$ such that $\{v\} \cup N[w] \subseteq V(H)$.
\end{lemma}
\begin{proof}
We construct a hole $H$.
Starting at $w$, perform two traversals of the cycle $K$ by moving in the two directions
of the cycle. Each traversal is stopped when reaching a vertex that is adjacent to $v$.
The hole $H$ is the subgraph of $G$ induced by $v$ and the vertices visited by the two traversals.
\end{proof}

\subsection{The inter-clique graph}

\begin{figure}
\centering
\subfigure[$G$]{\includegraphics{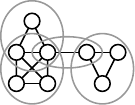}}%
\qquad%
\subfigure[$T_G$]{\includegraphics{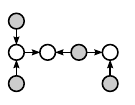}}%
\caption{A Ptolemaic graph $G$, and its inter-clique graph $T_G$.
In figure~(a), the gray ellipses denote the maximal cliques in $G$.
In figure~(b), the nodes of $T_G$ that correspond to maximal cliques in $G$
are colored gray.}\label{fig:inter-clique}
\end{figure}

The \emph{inter-clique} graph of an undirected graph $G$ is a directed graph $T_G$.
The vertex set $V(T_G)$ (whose elements are called \emph{nodes})
consists of all nonempty sets $B \subseteq V(G)$ such that $B$ is either
a maximal clique in $G$ or the intersection of a set of maximal cliques in $G$.
For two nodes $B_1,B_2 \in V(T_G)$, there is a directed edge $\edge{B_1}{B_2}$ in $T_G$
if and only if $B_2 \subset B_1$ and there is no node $B \in V(T_G)$ such that $B_2 \subset B \subset B_1$.
See Figure~\ref{fig:inter-clique}.
For a node $B \in V(T_G)$, we define the set $\Delta_G(B)$ as
\[ 
\Delta_G(B) = B \setminus \bigcup \{ C : \text{$C$ is a maximal clique and $B \not\subseteq C$}\}.
\]

\begin{lemma}\label{lem:delta-C-1}
Let $G$ be a graph.
Let $B \in V(T_G)$ and $v \in \Delta(B)$.
Then, a maximal clique $C$ in $G$ contains $v$ if and only if $B \subseteq C$.
\end{lemma}
\begin{proof}
Let $C$ be a maximal clique in $G$.

($\Rightarrow$)
Suppose that $v \in C$.
We want to show that $B \subseteq C$.
Suppose for contradiction that $B \not\subseteq C$.
By the definition of $\Delta(B)$, the assumption $B \not\subseteq C$ implies that
$\Delta(B)$ is disjoint from $C$.
Since $v \in \Delta(B)$, we have that $v \notin C$, which is a contradiction to the hypothesis that $v \in C$.
Therefore, $B \subseteq C$.

($\Leftarrow$)
Conversely, suppose that $B \subseteq C$.
Then, $\Delta(B) \subseteq B \subseteq C$, and therefore $v \in C$.
\end{proof}

\begin{lemma}\label{lem:delta-C-alt}
Let $G$ be a graph.
For every node $B \in V(T_G)$,
$\Delta(B) = B \setminus \bigcup_{\edge{B}{B'} \in E(T_G)} B'$.
\end{lemma}
\begin{proof}
Let $\edge{B}{B_1},\ldots,\edge{B}{B_d}$ be all the edges in $T_G$ that exit $B$.
This means $B_i \subset B$ for all $i$.
We need to show that $\Delta(B) = B \setminus \bigcup_{i=1}^d B_i$.

To show that $\Delta(B) \subseteq B \setminus \bigcup_{i=1}^d B_i$,
suppose for contradiction that there is a vertex $v \in \Delta(B)$ such that 
$v \notin B \setminus \bigcup_{i=1}^d B_i$.
Since $v \in B$, it must be that $v \in \bigcup_{i=1}^d B_i$.
Let $i$ be an index such that $v \in B_i$.
Since $B_i$ is a node of $T_G$, there are maximal cliques $C_1,\ldots,C_s$
such that $B_i = \bigcap_{j=1}^s C_j$ (where $s \geq 1$).
For every $j \leq s$, $v \in B_i \subseteq C_j$, and by Lemma~\ref{lem:delta-C-1}
we conclude that $B \subseteq C_j$.
Since $B \subseteq C_j$ for every $j \leq s$, we have that
$B \subseteq \bigcap_{j=1}^s C_j = B_i$, which is a contradiction.
Therefore, $\Delta(B) \subseteq B \setminus \bigcup_{i=1}^d B_i$.

To show that $\Delta(B) \supseteq B \setminus \bigcup_{i=1}^d B_i$,
suppose for contradiction that there is a vertex $v \in B \setminus \bigcup_{i=1}^d B_i$ such that 
$v \notin \Delta(B)$.
By the definition of $\Delta(B)$, the fact that $v \notin \Delta(B)$ implies that
there is a maximal clique $C$ that contains $v$ and $B \not\subseteq C$.
Since $B \cap C \neq \emptyset$ and $B \cap C$ is the intersection of maximal cliques,
the graph $T_G$ contains a node $B \cap C$.
Since $B \not\subseteq C$, we have that $B \cap C \subset B$.
By the definition of $T_G$, there is a path in $T_G$ from $B$ to $B \cap C$.
Let $B'$ be the node immediately following $B$ on this path.
By the definition of $B_1,\ldots,B_d$, we have that $B' = B_i$ for some index $i$.
We have that $v \in B \cap C \subseteq B_i$, which is a contradiction to the assumption that
$v \in B \setminus \bigcup_{i=1}^d B_i$.
Therefore, $\Delta(B) \supseteq B \setminus \bigcup_{i=1}^d B_i$.
Combining with the previous inclusion, we conclude that
$\Delta(B) = B \setminus \bigcup_{i=1}^d B_i$.
\end{proof}

\begin{lemma}\label{lem:delta-C-0}
Let $G$ be a graph.
Then, $\{\Delta(B) : B \in V(T_G)\}$ is a partition of $V(G)$.
\end{lemma}
\begin{proof}
We need to show that every vertex $v \in V(G)$ belongs to a unique set $\Delta(B)$.

Consider a vertex $v \in V(G)$.
We first show that there is a node $B_v \in V(T_G)$ such that $v \in \Delta(B_v)$.
Let $\C_v$ be the set of all maximal cliques that contain $v$.
Note that $\C_v \neq \emptyset$.
Let $B_v = \bigcap_{C \in \C_v} C$.
Clearly, $v \in B_v$.
$B_v$ is a nonempty intersection of maximal cliques, so $B_v$ is a node of $T_G$.
By the definition of $B_v$,
every maximal clique $C$ such that $B_v \not\subseteq C$ satisfies $C \notin \C_v$ and thus $v \notin C$.
By the definition of $\Delta(B_v)$, we conclude that $v \in \Delta(B_v)$.

We now show that if $v \in \Delta(B)$ for some node $B \in V(T_G)$, then $B = B_v$.

($B \supseteq B_v$)
$B$ is a node of $T_G$, and thus $B$ is the intersection of maximal cliques $C_1,\ldots,C_s$.
Let $\C = \{C_1,\ldots,C_s\}$.
Since $v \in B$, we have $v \in C_i$ for every $i \leq s$.
Thus, every $C_i$ belongs to the set $\C_v$.
Since $\C \subseteq \C_v$, taking the intersection yields the inclusion
\[ B = \bigcap_{C_i \in \C} C_i \supseteq \bigcap_{C \in \C_v} C = B_v. \]
Therefore, $B \supseteq B_v$.

($B \subseteq B_v$)
Consider any maximal clique $C \in \C_v$.
Since $v \in \Delta(B)$ and $v \in C$,
Lemma~\ref{lem:delta-C-1} implies that $B \subseteq C$.
Since this holds for every $C \in \C_v$, we have
$B \subseteq \bigcap_{C \in \C_v} C = B_v$.

From the two inclusions $B \supseteq B_v$ and $B \subseteq B_v$ it follows that $B = B_v$.
Therefore, every vertex $v \in V(G)$ belongs to a unique set $\Delta(B_v)$.
\end{proof}

\begin{lemma}\label{lem:delta-C-2}
Let $G$ be a graph.
For every node $B \in V(T_G)$, the set $\Delta(B)$ is a module in $G$.
\end{lemma}
\begin{proof}
Let $B$ be a node of $T_G$.
If $\Delta(B) = \emptyset$, then $\Delta(B)$ is a trivial module.
Thus, suppose that $\Delta(B) \neq \emptyset$.

We first show that for every $v \in \Delta(B)$, the closed neighborhood $N[v]$
is exactly the union of all maximal cliques that contain $B$.
Let
\[ \C_B = \{ C : \text{$C$ is a maximal clique in $G$ and $B \subseteq C$} \}.\]
Fix $v \in \Delta(B)$.

($N[v] \supseteq \bigcup \C_B$)
Let $u$ be a vertex in $\bigcup \C_B$.
Let $C$ be a maximal clique in $\C_B$ that contains $u$.
By the definition of $\C_B$, $B \subseteq C$.
Since $v \in \Delta(B)$ and $B \subseteq C$,
Lemma~\ref{lem:delta-C-1} implies $v \in C$.
Since $u,v \in C$ and $C$ is a clique, $u \in N[v]$.

($N[v] \subseteq \bigcup \C_B$)
Let $u$ be a vertex in $N[v]$.
Since $\{u,v\}$ is a clique, there is a maximal clique $C$ that contains $u$ and $v$.
Since $v \in \Delta(B)$ and $v \in C$,
Lemma~\ref{lem:delta-C-1} implies that $B \subseteq C$, and therefore $C \in \C_B$.
Thus, $u \in C \subseteq \bigcup \C_B$.

We have shown that for every $v \in \Delta(B)$, $N[v] = \bigcup \C_B$.
Therefore, for every $v \in \Delta(B)$,
\[ N(v) \setminus \Delta(B) = N[v] \setminus \Delta(B) = \bigcup \C_B \setminus \Delta(B).\]
Thus, for every $u,v \in \Delta(B)$,
$N(u) \setminus \Delta(B) = N(v) \setminus \Delta(B)$.
\end{proof}

\begin{theorem}[\citet{uehara2009laminar}]\label{thm:ptolemaic-2}
A graph $G$ is Ptolemaic if and only if $\und{T_G}$ is acyclic.
Additionally, for a given Ptolemaic graph $G$, the graph $T_G$ can
be constructed in polynomial time.
\end{theorem}

\subsection{The inter-clique graph of a Ptolemaic graph}

This section presents properties of the inter-clique graphs of Ptolemaic graphs.
For the remainder of the section, we assume that $G$ is a Ptolemaic graph.
Let $\C(G)$ denote the set of all maximal cliques in $G$.

\begin{lemma}[\citet{agrawal2021polynomial}]\label{lem:TG-1}
Let $B \in V(T_G)$ be a node.
Then $B$ is a maximal clique in $G$ if and only if there is no edge that enters $B$ in $T_G$.
\end{lemma}

\begin{lemma}\label{lem:C-Bi}
Let $C$ be a maximal clique in $G$ and let $B_1,\ldots,B_d$ be the neighbors of $C$ in $\und{T_G}$.
Then,
\begin{enumerate}
\item
$\Delta(C),B_1,\ldots,B_d$ is a partition of $C$.
\item
For every vertex $v \in V(G) \setminus C$, $v$ is non-adjacent to $\Delta(C)$.
\item
For every vertex $v \in V(G) \setminus C$, $v$ is adjacent to at most one set from the sets $B_1,\ldots,B_d$.
\end{enumerate}
\end{lemma}
\begin{proof}
(1)
By Lemma~\ref{lem:TG-1}, $\edge{C}{B_i} \in E(T_G)$ for every $i$,
and therefore $B_i \subset C$ for every $i$.
By Lemma~\ref{lem:delta-C-alt}, $\Delta(C) = C \setminus \bigcup_{i=1}^d B_i$.
Thus, to prove the first part of the lemma, it remains to show that
the sets $B_1,\ldots,B_d$ are pairwise disjoint.
Suppose for contradiction that there are indices $i \neq j$ such that $B_i \cap B_j \neq \emptyset$.
We have $B_i \not\subset B_j$ (otherwise $B_i \subset B_j \subset C$, and therefore
$\edge{C}{B_i} \notin E(T_G)$, which is a contradiction).
Since $B_i \cap B_j \neq \emptyset$, we have that $B' = B_i \cap B_j$ is a node of $T_G$.
Since $B' \subset B_i$ and $B' \subset B_j$, $T_G$ contains a path from $B_i$ to $B'$
and a path from $B_j$ to $B'$.
These paths do not pass through $C$ since there is no edge that enters $C$.
The existence of these two paths, along with the edges $\edge{C}{B_i}$ and $\edge{C}{B_j}$ implies that
the underlying graph $\und{T_G}$ contains a cycle.
This is a contradiction to Theorem~\ref{thm:ptolemaic-2}.
Thus, the sets $B_1,\ldots,B_d$ are pairwise disjoint.

(2)
Suppose for contradiction that $v \in V(G) \setminus C$ is adjacent to a vertex $u \in \Delta(C)$.
Since $\{u,v\}$ is a clique, there is a maximal clique $C'$ such that $\{u,v\} \subseteq C'$.
Because $v \in V(G) \setminus C$ and $v \in C'$, we have $C' \neq C$.
Since $u \in \Delta(C)$ and $u \in C'$,
Lemma~\ref{lem:delta-C-1} implies that $C \subseteq C'$.
However, because both $C$ and $C'$ are maximal cliques, the inclusion $C \subseteq C'$ implies that
$C = C'$, which is a contradiction.

(3)
Suppose for contradiction that a vertex $v \in V(G) \setminus C$ is adjacent to a vertex $u_1 \in B_i$ and
to a vertex $u_2 \in B_j$, where $i \neq j$.
The set $\{v,u_1,u_2\}$ is a clique, and thus there is a maximal clique $C'$ such that
$\{v,u_1,u_2\} \subseteq C'$.
We have that $C' \neq C$ since $v \in C'$ and $v \notin C$.
Since $u_1 \in B_i \subset C$ and $u_1 \in C'$, we have that $C \cap C' \neq \emptyset$.
Therefore, $C \cap C'$ is a node in $T_G$.
In $T_G$, there is a path from $C$ to $C \cap C'$ (since $C \cap C' \subset C$),
and let $B$ be the node immediately following $C$ on this path.
Since $\edge{C}{B_1},\ldots, \edge{C}{B_d}$ are all the edges that exit $C$,
we have that $B = B_s$ for some index $s$.
Since there is a path from $B_s$ to $C \cap C'$, we have $C \cap C' \subseteq B_s$.
Since $u_1 \in C \cap C'$, we have $u_1 \in B_s$.
Thus, $u_1 \in B_i \cap B_s$, so $B_i \cap B_s \neq \emptyset$.
Since $B_1,\ldots,B_d$ are pairwise disjoint (by the first part of the lemma),
this non-empty intersection forces $i = s$.
Similarly, $u_2 \in B_j$ and $u_2 \in C \cap C' \subseteq B_s$.
Therefore, $j = s$.
We conclude that $i = j$, which is a contradiction.
\end{proof}

\begin{figure}
\centering
\subfigure[$T_G$]{\includegraphics{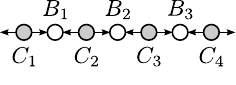}}%
\qquad%
\subfigure[$G$]{\includegraphics{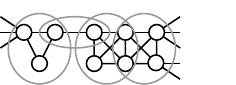}}%
\caption{A path $C_1,B_1,C_2,B_2,\ldots,C_4$ in $\und{T_G}$ such that
the nodes in the path have degree~2 in $\und{T_G}$, and the corresponding vertices of $G$.
The nodes $C_1,\ldots,C_4$ are maximal cliques in $G$.\label{fig:path}}
\end{figure}

The following lemma gives properties of paths in $\und{T_G}$ in which all nodes have degree~2
in $\und{T_G}$. See Figure~\ref{fig:path} for an example.
\begin{lemma}\label{lem:TG-4}
Let $C_1,B_1,C_2,B_2,\ldots,C_{t-1},B_{t-1},C_t$ be a path in $\und{T_G}$ such that
the nodes of the path have degree~2 in $\und{T_G}$ and $C_1$ is a maximal clique in $G$.
Then,
\begin{enumerate}
\item\label{enu:TG-cliques}
The nodes $C_1,\ldots,C_t$ are maximal cliques in $G$, and
the nodes $B_1,\ldots,B_{t-1}$ are not maximal cliques in $G$.

\item\label{enu:TG-CiCj}
For every $i$ and $j$,
$C_i \cap C_j \neq \emptyset$ if and only if $|i-j| \leq 1$.

\item\label{enu:TG-C}
For every $1 < i < t$ and every $C \in \C(G) \setminus \{C_{i-1},C_i,C_{i+1}\}$,
$C_i \cap C = \emptyset$.

\item\label{enu:TG-Bi-1}
For every $1 < i < t$,
$\Delta(C_i) = C_i \setminus (B_{i-1} \cup B_i)$.

\item\label{enu:TG-Bi-2}
For every $i < t$, $\Delta(B_i) = B_i = C_i \cap C_{i+1}$.

\item\label{enu:TG-Nv-1}
If $v \in \Delta(C_i)$, then $N[v] = C_i$.

\item\label{enu:TG-Nv-2}
If $v \in B_i$, then $N[v] = C_i \cup C_{i+1}$.

\item\label{enu:TG-dist}
If $u \in C_i$, $v \in C_j$, and $i \leq j$, then
the distance between $u$ and $v$ in the graph $G[C_1 \cup \cdots \cup C_t]$
is at least $j-i-1$.
\end{enumerate}
\end{lemma}
\begin{proof}
Properties~\ref{enu:TG-cliques}, \ref{enu:TG-CiCj}, and~\ref{enu:TG-C} are proved
in~\citep[Observation~31 and Lemma~32]{agrawal2021polynomial}.
Properties~\ref{enu:TG-Bi-1} and~\ref{enu:TG-Nv-1} follow from Lemma~\ref{lem:C-Bi}.
We now prove that properties~\ref{enu:TG-Bi-2} and~\ref{enu:TG-Nv-2} hold.

(\ref{enu:TG-Bi-2})
Let $i < t$.
By Lemma~\ref{lem:TG-1}, the two edges incident to $B_i$ in $T_G$ are $\edge{C_i}{B_i}$
and $\edge{C_{i+1}}{B_i}$, and thus no edge exits $B_i$.
By property~\ref{enu:TG-CiCj}, $C_i \cap C_{i+1} \neq \emptyset$.
By the definition of $T_G$, $B = C_i \cap C_{i+1}$ is a node of $T_G$.
Since $B \subset C_i$ and $B \subset C_{i+1}$,
there are paths in $T_G$ from $C_i$ to $B$ and from $C_{i+1}$ to $B$.
Suppose for contradiction that $B \neq B_i$.
Then the edges $\edge{C_i}{B_i}$ and $\edge{C_{i+1}}{B_i}$
combined with the paths from $C_i$ and $C_{i+1}$ to $B$,
form a cycle in $\und{T_G}$.
This is a contradiction to Theorem~\ref{thm:ptolemaic-2}.
Therefore, $B_i = B = C_i \cap C_{i+1}$.

Since no edge exits $B_i$, Lemma~\ref{lem:delta-C-alt} implies that $\Delta(B_i) = B_i$.

(\ref{enu:TG-Nv-2})
Let $v \in B_i$.
By the proof of Lemma~\ref{lem:delta-C-2}, $N[v] = \bigcup \C_{B_i}$,
where
\[ \C_{B_i} = \{ C : \text{$C$ is a maximal clique in $G$ and $B_i \subseteq C$} \}.\]
By properties~\ref{enu:TG-C} and~\ref{enu:TG-Bi-2}, $\C_{B_i} = \{C_i,C_{i+1}\}$.

(\ref{enu:TG-dist})
We prove this property by induction on $j-i$.
The base case, when $j-i \leq 1$, is trivial.
Assume now that $j-i > 1$.
Let $P$ be a shortest path from $u$ to $v$ in $G[C_1 \cup \cdots \cup C_t]$.
Let $u'$ be the first vertex on $P$ that does not belong to the set $C_1 \cup \cdots \cup C_i$.
By properties~\ref{enu:TG-Bi-1} and~\ref{enu:TG-Bi-2},
we have $u' \in C_{i+1}$.
By the induction hypothesis, the subpath of $P$ from $u'$ to $v$ has
length at least $j-(i+1)-1$.
Therefore, the length of $P$ is at least $j-i-1$.
\end{proof}

\begin{lemma}\label{lem:P-replace}
Let $C_1,B_1,C_2,B_2,C_3,B_3,C_4$ be a path in $\und{T_G}$ such that
the nodes in the path have degree~2, and $C_1,\ldots,C_4$ are maximal cliques in $G$.
Let $G'$ be a graph that is obtained from $G$ by deleting
the vertices in $(C_2 \cup C_3) \setminus (B_1 \cup B_3)$ and adding an edge between
every vertex in $B_1$ and every vertex in $B_3$.
Then, $T_{G'}$ is the graph obtained from $T_G$ by deleting the nodes $C_2,C_3,B_2$,
adding a new node $C_0 = B_1 \cup B_3$, and adding edges $\edge{C_0}{B_1}$ and $\edge{C_0}{B_3}$.
\end{lemma}
\begin{proof}
We first show that $V(T_{G'}) = V(T_G) \setminus \{C_2,C_3,B_2\} \cup \{C_0\}$.
We achieve this by first showing the relation between $\C(G)$ and $\C(G')$,
and then showing the relation between $V(T_G) \setminus \C(G)$ and $V(T_{G'}) \setminus \C(G')$.

We first show that $\C(G') = \C'$,
where $\C' = \C(G) \setminus \{C_2,C_3\} \cup \{C_0\}$.

($\supseteq$)
Let $C \in \C'$.
We need to show that $C$ is a maximal clique in $G'$.

\textbf{Case 1:} $C = C_0$.\\
By the definition of $G'$, $C_0$ is a clique in $G'$.
Suppose for contradiction that $C_0$ is not a maximal clique in $G'$.
Thus, there is a vertex $w \in V(G') \setminus C_0$
that is adjacent to all vertices of $C_0$ in $G'$.
Since every edge in $E(G') \setminus E(G)$ has both endpoints in $C_0$,
no edge in $E(G') \setminus E(G)$ is incident to $w$.
Thus, $w$ is adjacent to all vertices of $C_0$ in the original graph $G$.
Let $v_1 \in B_1$.
Since $v_1 \in B_1 \subseteq C_0$, $w$ is adjacent to $v_1$, so $w \in N[v_1]$.
By Lemma~\ref{lem:TG-4}, $w \in N[v_1] = C_1 \cup C_2$.
Similarly, let $v_3 \in B_3$.
Since $v_3 \in B_3 \subseteq C_0$, we have $w \in N[v_3] = C_3 \cup C_4$.
Combining the statements above, $w$ must be in the intersection
$(C_1 \cup C_2) \cap (C_3 \cup C_4)$.
By Lemma~\ref{lem:TG-4}, $w \in (C_1 \cup C_2) \cap (C_3 \cup C_4) = B_2$.
This means that $w$ is deleted when forming $G'$, which is
a contradiction to the assumption that $w \in V(G')$.
Therefore, $C_0$ is a maximal clique in $G'$.

\textbf{Case 2:} $C \neq C_0$.\\
By the definition of $\C'$, $C \in \C(G) \setminus \{C_2,C_3\}$,
so $C$ is a maximal clique in $G$.
The vertices deleted from $G$ to form $G'$ are
$\Vd = (C_2 \cup C_3) \setminus (B_1 \cup B_3)$.
By Lemma~\ref{lem:TG-4}, the only cliques in $\C(G) \setminus \{C_2,C_3\}$ that intersect $C_2 \cup C_3$
are $C_1$ and $C_4$.
If $C \notin \{C_1, C_4\}$, then $C \cap \Vd = \emptyset$, meaning $C \subseteq V(G')$.
If $C = C_1$, then $C_1 \cap \Vd = \emptyset$ as $C_1 \cap (C_2 \cup C_3) = B_1$.
Thus, $C \subseteq V(G')$.
Similarly, if $C = C_4$, then $C \subseteq V(G')$.
Therefore, $C \subseteq V(G')$ in all cases.
The graph $G'$ differs from $G[V(G')]$ by the new edges between $B_1$ and $B_3$.
Therefore, $C$ is a clique in $G'$.
Suppose for contradiction that $C$ is not a maximal clique in $G'$.
Thus, there is a vertex $w \in V(G') \setminus C$
that is adjacent to all vertices of $C$ in $G'$.

We claim that $w$ is also adjacent to all vertices of $C$ in $G$.
Suppose for contradiction that there is a vertex $u \in C$ such that $w$ is not adjacent to $u$ in $G$.
Thus, $\edge{u}{w} \in E(G') \setminus E(G)$.
By the definition of $G'$, one endpoint of $\edge{u}{w}$ is in $B_1$ and the other in $B_3$.
Without loss of generality, let $u \in B_1$ and $w \in B_3$.
By Lemma~\ref{lem:TG-4}, the maximal cliques in $G$ that contain $u$ are $C_1$ and $C_2$.
Since $C \in \C(G) \setminus \{C_2,C_3\}$, we obtain that $C = C_1$.
By Lemma~\ref{lem:TG-4}, $w$ is not adjacent to $C$ in $G$.
By the definition of $G'$, the edges in $E(G') \setminus E(G)$ are between vertices in $B_1$ and
vertices in $B_3$.
Therefore, $w$ is not adjacent to $C \setminus B_1$ in $G'$,
which is a contradiction.
Thus, $w$ is adjacent to all vertices of $C$ in $G$.
Therefore, $C \cup \{w\}$ is a clique in $G$, which is a contradiction to the fact
that $C$ is a maximal clique in $G$.
Therefore, $C$ is a maximal clique in $G'$.
This completes the proof that $\C(G') \supseteq \C'$.

($\subseteq$)
Suppose for contradiction that $\C(G') \not\subseteq \C'$.
Let $C \in \C(G') \setminus \C'$. By definition, $C \neq C_0$.
We claim that $C$ intersects at most one set from $B_1,B_3$.
Suppose for contradiction that $C$ intersects both $B_1$ and $B_3$.
Since $C_0$ is a clique in $G'$ and $C \neq C_0$ is a maximal clique in $G'$,
we must have $C \not\subseteq C_0$. Let $x \in C \setminus C_0$.
Since $C$ is a clique in $G'$ that intersects $B_1$ and $B_3$,
$x$ is adjacent to both $B_1$ and $B_3$ in $G'$.
Since $x \notin C_0$ and the edges in $E(G') \setminus E(G)$ are contained in $C_0$,
no edge in $E(G') \setminus E(G)$ is incident to $x$.
Thus, $x$ is adjacent to both $B_1$ and $B_3$ in $G$.
By Lemma~\ref{lem:TG-4}, the only vertices in $G$ adjacent to both $B_1$ and $B_3$ are
the vertices in $B_2$.
Thus $x \in B_2$.
However, $B_2$ is disjoint from $V(G')$,
which contradicts the fact that $x \in C \subseteq V(G')$.
Therefore, $C$ intersects at most one set from $B_1,B_3$.
Consequently, there is no edge in $E(G') \setminus E(G)$ that is contained in $C$.
Thus, $C$ is a clique in $G$.

Since $C \notin \C'$ and $C \notin \{C_2,C_3\}$, we have $C \notin \C(G)$.
Therefore, there is a maximal clique $C' \in \C(G)$ such that $C \subset C'$.
Since $C$ is a maximal clique in $G'$, $C'$ must not be a clique in $G'$
(otherwise $C \subset C'$ contradicts the maximality of $C$).
This means that $C'$ must contain a vertex from
$\Vd = (C_2 \setminus B_1) \cup (C_3 \setminus B_3)$.
Without loss of generality, suppose that $C'$ contains a vertex $v \in C_2 \setminus B_1$.
By Lemma~\ref{lem:TG-4}, $N[v] \subseteq C_2 \cup C_3$.
Since $C'$ is a clique containing $v$, $C' \subseteq C_2 \cup C_3$.
Therefore, $C \subseteq C' \cap V(G') \subseteq (C_2 \cup C_3) \cap V(G') = C_0$.
Since $C$ and $C_0$ are maximal cliques in $G'$, this forces $C = C_0$, which is a contradiction.
Thus, $\C(G') \subseteq \C'$.

We showed above that $\C(G') = \C'$.
We next show that $V(T_{G'}) \setminus \C(G') = (V(T_G) \setminus \C(G)) \setminus \{B_2\}$.
Recall that $V(T_G) \setminus \C(G)$ and $V(T_{G'}) \setminus \C(G')$ contain
the nonempty intersections of at least two cliques from $\C(G)$ and from $\C(G')$, respectively.
We therefore consider intersections of cliques from $\C(G)$ that involve
the deleted cliques $\C(G) \setminus \C(G') = \{C_2,C_3\}$ and
intersections of cliques from $\C(G')$ that involve the new cliques
$\C(G') \setminus \C(G) = \{C_0\}$.
The maximal cliques in $G$ that intersect $C_2$ are $C_1$ and $C_3$.
The maximal cliques in $G$ that intersect $C_3$ are $C_2$ and $C_4$.
Therefore, the nonempty intersections of cliques from $\C(G)$ that contain at least one clique
from $\{C_2,C_3\}$ are
$C_1 \cap C_2 = B_1$, $C_2 \cap C_3 = B_2$, and $C_3 \cap C_4 = B_3$.
The maximal cliques in $G'$ that intersect $C_0$ are $C_1$ and $C_4$.
Therefore, the nonempty intersections of cliques from $\C(G')$ that contain $C_0$ are
$C_1 \cap C_0 = B_1$ and $C_4 \cap C_0 = B_3$.
Therefore,
$V(T_{G'}) \setminus \C(G') = (V(T_G) \setminus \C(G)) \setminus \{B_2\}$.

From the above, we obtain that
$V(T_{G'}) = V(T_G) \setminus \{C_2,C_3,B_2\} \cup \{C_0\}$.
Finally, we determine the relationship between $E(T_G)$ and $E(T_{G'})$.
The edges of $T_G$ that are incident on the deleted nodes $C_2,C_3,B_2$ are
removed.
For the added edges, consider the new node $C_0$ and its relationship to the
remaining nodes in $T_{G'}$.
The nodes of $T_{G'}$ that are proper subsets of $C_0$ are exactly $B_1$ and $B_3$.
It follows that
$E(T_{G'}) = E(T_G) \setminus \{\edge{C_2}{B_1}, \edge{C_2}{B_2}, \edge{C_3}{B_2}, \edge{C_3}{B_3}\}
\cup \{\edge{C_0}{B_1}, \edge{C_0}{B_3}\}$.
\end{proof}

\begin{lemma}\label{lem:path-1}
Let $G_1$ be a graph, and let $A$ be a deletion set of $G_1$.
Let $C_1,B_1,C_2,B_2,C_3$ be a path in $\und{T_{G_1-A}}$ such that
the nodes of the path have degree~2, $C_1,C_2,C_3$ are maximal cliques in $G_1-A$,
and $C_1,C_2,C_3$ are not adjacent to $A$ in $G_1$.
If $H$ is a hole in $G_1$ and $V(H) \cap C_2 \neq \emptyset$, then
$|V(H) \cap B_1| = 1$, $|V(H) \cap B_2| = 1$, and
$V(H) \cap (C_2 \setminus (B_1 \cup B_2)) = \emptyset$.
\end{lemma}
\begin{proof}
Let $v \in V(H) \cap C_2$.
For every $u \in C_2 \setminus (B_1 \cup B_2)$, Lemma~\ref{lem:TG-4} implies that
$N_{G_1}[u] = N_{G_1-A}[u] = C_2$.
Since $C_2$ is a clique, every two neighbors of $u$ are adjacent.
Thus, $u$ cannot be in $H$, otherwise the edge between the two neighbors of $u$ in $H$
is a chord in $H$.
Therefore, $V(H) \cap (C_2 \setminus (B_1 \cup B_2)) = \emptyset$.
This means $v \in B_1 \cup B_2$.
Without loss of generality, suppose that $v \in B_1$.

Let $v_1,v_2$ be the two neighbors of $v$ in $H$.
By Lemma~\ref{lem:TG-4}, $N_{G_1}[u] = N_{G_1-A}[u] = C_1 \cup C_2$ for every $u \in B_1$.
In particular, $N_{G_1}[v] = C_1 \cup C_2$, so $v_1,v_2 \in C_1 \cup C_2$.

If there is a vertex $w \in V(H) \cap B_1$ with $w \neq v$, then
$v$ and $w$ are adjacent since $B_1$ is a clique.
Thus, $w \in \{v_1,v_2\}$.
Without loss of generality, assume that $w = v_1$.
Since $N_{G_1}[v] = N_{G_1}[w] = C_1 \cup C_2$ and $\edge{v}{v_2} \in E(G_1)$,
we have that $\edge{v_1}{v_2} \in E(G_1)$.
Therefore, $\edge{v_1}{v_2}$ is a chord in $H$, which is a contradiction.
Therefore, $V(H) \cap B_1 = \{v\}$.

Since $C_1$ is a clique, the vertices $v_1$ and $v_2$ cannot both be in $C_1$.
Without loss of generality, assume that $v_1 \notin C_1$.
Thus, $v_1 \in C_2 \setminus C_1$.
Since $V(H) \cap (C_2 \setminus (B_1 \cup B_2)) = \emptyset$,
we have that $v_1 \in B_2$.
By the same argument used for $B_1$, we obtain that $V(H) \cap B_2 = \{v_1\}$.
\end{proof}

\begin{lemma}\label{lem:path-2}
Let $G_1$ be a graph, and let $A$ be a deletion set of $G_1$.
Let $R \subseteq V(G_1) \setminus A$.
Let $C_1,B_1,C_2,B_2,C_3$ be a path in $\und{T_{G_1-A-R}}$ such that
the nodes of the path have degree~2, $C_1,C_2,C_3$ are maximal cliques in $G_1-A-R$,
and $C_1$, $C_2$, and $C_3$ are not adjacent to $A \cup (R \cap N_{G_1}(A))$ in $G_1$.
If $H$ is an obstruction in $G_1$ and $V(H) \cap C_2 \neq \emptyset$, then
$H$ is a hole.
\end{lemma}
\begin{proof}
Suppose for contradiction that $H$ is not a hole.
This means that $H$ is a gem.
Let $u$ be the center vertex of $H$. Note that $u$ is adjacent to every other vertex in $V(H)$.
If $u \in C_2$, then $V(H) \subseteq N[u] \subseteq V(G_1) \setminus A$
(the second inclusion follows from the conditions of the lemma).
Therefore, $H$ is an obstruction in $G_1-A$, which is a contradiction.
Thus, $u \notin C_2$.
Let $v$ be a vertex in $V(H) \cap C_2$.
Since $u,v$ are adjacent, $u \in N(v)$.
By Lemma~\ref{lem:TG-4}, $N(v) \subseteq C_1 \cup C_2 \cup C_3 \cup (R \setminus N(A))$.
Since no vertex in $C_1 \cup C_2 \cup C_3 \cup (R \setminus N(A))$ is adjacent to $A$,
we have that $u$ is not adjacent to $A$.
Therefore, $V(H) \subseteq N[u] \subseteq V(G_1) \setminus A$, which is a contradiction.
\end{proof}

\subsection{Hitting sets}

A set $S$ is a \emph{hitting set} of a collection of sets $\D$ if
$D \cap S \neq \emptyset$ for every set $D \in \D$.
In the \textsc{$d$-Hitting Set} problem, the input consists of a collection of sets $\D$,
where each set in $\D$ has a size of at most $d$, and an integer $k$.
The goal is to decide whether there is a hitting set $S$ of $\D$ such that $|S| \leq k$.

\begin{theorem}[\citet{fomin2013polynomial}]\label{thm:hitting-set}
Let $d$ be a fixed positive integer.
There is a polynomial time algorithm that, given an instance $(\D,k)$ of \textsc{$d$-Hitting Set} with $k \geq 1$,
computes a set $\D' \subseteq \D$ such that
\begin{enumerate}
\item
$|\D'| = O(d! k^d)$.
\item
If $S$ is a hitting set of $\D'$ of size at most $k$, then $S$ is a hitting set of $\D$.
\end{enumerate}
\end{theorem}

\section{The kernel}

\subsection{Overview}

Let $(G_0,k)$ be an instance of \textsc{Ptolemaic Vertex Deletion}.
If $k < 0$, the algorithm returns a fixed no-instance.
If $k = 0$, it tests whether $G_0$ is Ptolemaic and returns a fixed instance with the same answer.
If $k \geq |V(G_0)|$, it returns a fixed yes-instance.
Thus, we may assume that $1 \leq k < |V(G_0)|$.
The kernelization algorithm consists of the following steps.

\paragraph{Step 1}
First, the algorithm uses the approximation algorithm of Theorem~\ref{thm:approximation}
to construct a deletion set $A$ for the graph $G_0$.
Since the approximation factor is 68,
if $|A| > 68k$, then the instance $(G_0,k)$ is determined to be a no-instance.
We proceed assuming that $|A| \leq 68k$.

\paragraph{Step 2}
The second step is taken from \citet{agrawal2021polynomial}.
This step has two goals.
\begin{enumerate}
\item
The first goal is to reduce the independent degrees of the vertices in $A$.
The \emph{independent degree} of a vertex $v$ in a graph $G$ with respect to a set $E_I \subseteq E(G)$,
denoted $d^I_{G,E_I}(v)$, is the maximum size of an independent set in $G[N_{G-E_I}(v)]$.
The algorithm aims for $d^I_{G,E_I}(v) = O(k^2)$ for every vertex $v \in A$,
where $E_I$ is a set of edges that is constructed in this step.

\item
The second goal is to reduce the number of leaves of the graph $\und{T_{G-A-R}}$ to $O(k^3)$,
where $R \subseteq V(G)$ is a set of $O(k^5)$ vertices that is constructed in this step.

\end{enumerate}

In order to achieve this,
the algorithm transforms the current instance $(G_0,k)$ to an equivalent instance $(G,k',E_M)$
of a new problem called \textsc{Augmented Ptolemaic Vertex Deletion},
which is defined as follows.
\begin{description}
\item[Input]
A graph $G$, an integer $k$, and a set $E_M$ of pairs of vertices of $G$.

\item[Output]
Whether there is a set $S$ of size at most $k$ such that $S$ is a deletion set of $G$
and a hitting set of $E_M$.
\end{description}

Whenever the current parameter $k'$ becomes negative, the algorithm returns a fixed no-instance.
If $k' = 0$, it returns a fixed yes-instance if $G$ is Ptolemaic and $E_M = \emptyset$,
and a fixed no-instance otherwise.
Thus, whenever the algorithm continues, $1 \leq k' \leq k$.

\paragraph{Step 3}
In this step, the algorithm reduces the number of nodes of $T_{G-A-R}$ to $O(k^4)$.
This goal is achieved through the application of three reduction rules.
The first two rules reduce the number of nodes $B$ in $V(T_{G-A-R})$ that are adjacent to
the set $A \cup (R \cap N(A))$ in $G$.
The third rule limits the lengths of paths in $\und{T_{G-A-R}}$ whose
nodes have degree~2 and are not adjacent to $A \cup (R \cap N(A))$ in $G$.

\paragraph{Step 4}

In this step, the algorithm reduces the number of vertices in $G$ to $O(k^5)$.
This is achieved by considering a partition of $V(G-A-R)$ into clique-modules based on $T_{G-A-R}$.
Since the previous step bounded the number of nodes in $T_{G-A-R}$ to $O(k^4)$,
the number of resulting clique-modules is also $O(k^4)$.
After a final reduction rule is applied, the vertices of $G$ are distributed into three categories.
\begin{enumerate}
\item
$O(k^5)$ special vertices (which include all vertices in $A$ and $R$).

\item
Large modules: $O(k^3)$ clique-modules, each containing $O(k^2)$ vertices,
excluding special vertices.

\item
Small modules: $O(k^4)$ clique-modules, each containing $O(k)$ vertices,
excluding special vertices.
\end{enumerate}

\paragraph{Step 5}
In the final step, the algorithm transforms the instance $(G,k',E_M)$
(an instance of \textsc{Augmented Ptolemaic Vertex Deletion})
back into an instance $(G',k')$ of \textsc{Ptolemaic Vertex Deletion}.
This reduction is taken from \citep{agrawal2021polynomial}.
The reduction handles the hitting set constraint imposed by $E_M$ as follows.
For every pair $\{x,y\} \in E_M$, the reduction adds
$k+2$ paths $x,a_i,b_i,y$ of length three, using two new vertices $a_i,b_i$ for each path.
All new vertices are distinct across all paths and pairs, and have no additional incident edges.
The size of the new graph $G'$ is therefore
$|V(G')| = |V(G)|+2(k+2)|E_M| = |V(G)|+O(k |E_M|)$.

The correctness of this reduction follows from the proof of
Theorem~1 in \citet{agrawal2021polynomial}.
Thus, $(G,k',E_M)$ is a yes-instance if and only if
$(G',k')$ is a yes-instance.

\subsection{Reducing the number of leaves of $\und{T_{G-A-R}}$}

The algorithm in the following lemma is used to reduce the number of leaves of $\und{T_{G-A-R}}$.

\begin{lemma}[\citet{agrawal2021polynomial}]\label{lem:augmented}
There is a polynomial time algorithm that
given an instance $(G_0,k)$ of \textsc{Ptolemaic Vertex Deletion}
and a deletion set $A$ of $G_0$ with $|A| \leq 68k$,
either concludes that $(G_0,k)$ is a no-instance or
constructs an equivalent instance $(G,k',E_M)$ of \textsc{Augmented Ptolemaic Vertex Deletion}
along with auxiliary sets $E_I \subseteq E(G)$ and $R \subseteq V(G) \setminus A$ with the following
properties (if vertices of $A$ are deleted during preprocessing, $A$ denotes its remaining subset):
\begin{enumerate}[label=P\arabic*., ref=P\arabic*]
\item\label{enu:aug-kp}\label{enu:aug-first}
$k' \leq k$.

\item\label{enu:aug-A}
$A$ is a deletion set of $G$.

\item\label{enu:aug-e}
$A$ is a hitting set of $E_I \cup E_M$.

\item\label{enu:aug-EI}
If $K$ is a hole in $G$ that contains an edge from $E_I$ then there is a pair $e \in E_M$
such that $e \subseteq V(K)$.

\item\label{enu:aug-EM}
$|E_M| = O(k^2)$.

\item\label{enu:aug-R-EI}
Every hole in $G$ that contains a vertex from $R$ contains an edge from $E_I$.

\item\label{enu:aug-R-N}
For every $v \in R$, $N(v)\cap V(G-A-R)$ is a clique.

\item\label{enu:aug-R-size}
$|R| = O(k^5)$.

\item\label{enu:aug-dI}\label{enu:aug-2nd-last}
For every $v \in A$, $d^I_{G,E_I}(v) = O(k^2)$.

\item\label{enu:aug-T}\label{enu:aug-last}
The number of leaves of $\und{T_{G-A-R}}$ is $O(k^3)$.
\end{enumerate}
\end{lemma}

Until the final marking procedure, we maintain properties~\ref{enu:aug-first}--\ref{enu:aug-last}.
Thus, we must show that each reduction rule applied before marking preserves these properties.

For a set of pairs of vertices $F$ and a vertex $v$, we write
$F-v = \{e \in F : v \notin e\}$.

\begin{lemma}\label{lem:del-v}
Consider a reduction rule that given an instance $(G,k',E_M)$,
sets $A \gets A \setminus \{v\}$ and $E_I \gets E_I-v$,
and returns the instance $(G-v,k'',E_M-v)$,
where $v \in V(G) \setminus R$ and $k'' \leq k'$.
Then, the rule preserves properties~\ref{enu:aug-first}--\ref{enu:aug-2nd-last}.
\end{lemma}
\begin{proof}
Let $G' = G-v$, $A' = A \setminus \{v\}$, $E'_I = E_I-v$, and $E'_M = E_M-v$.
We demonstrate below that the rule preserves properties~\ref{enu:aug-e}, \ref{enu:aug-EI} and~\ref{enu:aug-R-EI}.
It is easy to verify that the rule preserves the remaining properties.

(\ref{enu:aug-e})
Let $e \in E'_I \cup E'_M$.
By construction, $e$ is also an element in the original set $E_I \cup E_M$.
By property~\ref{enu:aug-e}, $e \cap A \neq \emptyset$.
Since $e$ is a pair of vertices in the new instance, $v \notin e$.
Therefore, $e \cap A' = e \cap A \neq \emptyset$.
Thus, $A'$ is a hitting set of $E'_I \cup E'_M$.

(\ref{enu:aug-EI})
Let $K$ be a hole in $G'$ that contains an edge $e' \in E'_I$.
Since $G'$ is an induced subgraph of $G$, $K$ is also
a hole in $G$ that contains $e' \in E_I$.
By property~\ref{enu:aug-EI}, there is a pair $e \in E_M$
such that $e \subseteq V(K)$.
Since $K$ is a hole in $G'$, we have $v \notin V(K)$, which implies $v \notin e$.
Therefore, $e \in E'_M$.

(\ref{enu:aug-R-EI})
Let $K$ be a hole in $G'$ that contains a vertex $r \in R$.
Since $K$ is a hole in $G$ that contains a vertex $r \in R$,
by property~\ref{enu:aug-R-EI}, $K$ contains an edge $e \in E_I$.
Since $K$ is a hole in $G'$, we have $v \notin V(K)$, which implies $v \notin e$.
Therefore, $e \in E'_I$.
\end{proof}

We consider the undirected inter-clique graph $T = \und{T_{G-A-R}}$.
By property~\ref{enu:aug-A}, $G-A$ is a Ptolemaic graph.
Since $G-A-R$ is an induced subgraph of $G-A$ and the Ptolemaic property is closed under induced subgraphs,
$G-A-R$ is also a Ptolemaic graph.
Consequently, by Theorem~\ref{thm:ptolemaic-2}, $T$ is acyclic.

We now define two subsets of $V(T)$:
$X_1$ is the set of nodes in $T$ whose degree is not 2.
$X_2$ is the set of nodes in $V(T) \setminus X_1$ that have a distance of at most 3 to a node in $X_1$.

\begin{lemma}\label{lem:X1X2}
$|X_1 \cup X_2| = O(k^3)$.
\end{lemma}
\begin{proof}
Let $p$ be the number of leaves of $T$.
By property~\ref{enu:aug-T}, we have $p = O(k^3)$.
Since $T$ is acyclic, the number of nodes with degree at least 3 is at most $p$.
Thus, $|X_1| \leq 2p = O(k^3)$.

The removal of $X_1$ from $T$, resulting in $T-X_1$, yields a collection of disjoint paths.
The number of these paths is at most $|X_1|$.
Each path contains at most six vertices of $X_2$ (three vertices near each endpoint of the path).
Therefore, $|X_2| \leq 6 |X_1| = O(k^3)$.
\end{proof}

\begin{lemma}\label{lem:T-A}

The number of nodes $B \in V(T)$ such that $\Delta(B)$ is adjacent to $A$ in $G-E_I$ is $O(k^3)$.
\end{lemma}
\begin{proof}
Let $v$ be an arbitrary vertex in $A$.
Let $W_v = N_{G-E_I}(v) \setminus (A \cup R)$,
and let $I_v$ be a maximum independent set in the graph $G[W_v]$.
Since $I_v$ is also an independent set in the graph $G[N_{G-E_I}(v)]$,
by property~\ref{enu:aug-dI}, $|I_v| \leq d^I_{G,E_I}(v) = O(k^2)$.

We define $X_v = \{ B \in V(T) : \Delta(B) \cap I_v \neq \emptyset\}$.
By definition, if $B \in X_v$, then $\Delta(B)$ contains at least one vertex of $I_v$.
By Lemma~\ref{lem:delta-C-0}, a vertex in $I_v$ is contained in exactly one set $\Delta(B)$,
which implies that $|X_v| \leq |I_v|$.
To show equality, consider a node $B \in X_v$.
Since $I_v$ is an independent set in $G-A-R$ and $\Delta(B)$ is a clique,
the set $\Delta(B)$ can contain at most one vertex of $I_v$.
It follows that $|X_v| = |I_v|$.

Let $B$ be a node of $T$ such that $\Delta(B)$ is adjacent to $v$ in $G-E_I$ and $B \notin X_1 \cup X_2 \cup X_v$.
Let $u$ be a vertex in $\Delta(B) \cap N_{G-E_I}(v)$.
Since $B \notin X_v$, the set $\Delta(B)$ does not intersect $I_v$.
Therefore, $u \notin I_v$.
Since $u \in W_v$, the set $I_v \cup \{u\}$ is contained in $W_v$.
Due to the maximality of the independent set $I_v$ in $G[W_v]$,
the set $I_v \cup \{u\}$ cannot be independent.
Thus, $u$ is adjacent to some vertex $u' \in I_v$.
Since $u' \in I_v$ and $B \notin X_v$, we have $u' \notin \Delta(B)$.
Let $B' \in X_v$ be the unique node such that $u' \in \Delta(B')$
($B'$ is unique by Lemma~\ref{lem:delta-C-0}).
We define this as a \emph{match} of $B$ to $B'$.

Let $P = C_1,B_1,C_2,B_2,\ldots,C_t$ be the longest path in $T-X_1$ that passes through $B$ and
whose endpoints are maximal cliques in $G-A-R$.
By the definition of $X_2$, we have $B \in V(P) \setminus \{C_1,C_t,B_1,B_{t-1}\}$.
If $B = C_i$ for some $i$, then by Lemma~\ref{lem:TG-4},
\[ u' \in N_{G-A-R}(u) \setminus \Delta(B) = C_i \setminus \Delta(C_i) = B_{i-1} \cup B_i
 = \Delta(B_{i-1}) \cup \Delta(B_i). \]
Since $u' \in \Delta(B')$, we have $B' \in \{B_{i-1},B_i\}$.
If $B = B_i$ for some $i$, then by Lemma~\ref{lem:TG-4},
\[ u' \in N_{G-A-R}(u) \setminus \Delta(B) = (C_i \cup C_{i+1}) \setminus B_i =
\Delta(B_{i-1}) \cup \Delta(C_i) \cup \Delta(C_{i+1}) \cup \Delta(B_{i+1}), \]
and thus $B' \in \{B_{i-1},C_i,C_{i+1},B_{i+1}\}$.
In both cases, the distance between $B$ and its matched node $B'$ in $T$ is at most 2.
Moreover, $B' \notin X_1$ (since $B \notin X_1 \cup X_2$ and the distance between $B$ and $B'$ is at most 2).

The graph $T-X_1$ is a collection of disjoint paths.
For any matched node $B' \in X_v \setminus X_1$, the nodes that are matched to $B'$
must lie on the path of $T-X_1$ that contains $B'$.
Since the distance between a node $B'$ and a node $B$ that is matched to $B'$ is at most 2,
the number of nodes that are matched to $B'$ is at most 4.
Therefore, the number of matched nodes (nodes $B \in V(T)$ such that $\Delta(B)$ is adjacent to $v$ in $G-E_I$ and
$B \notin X_1 \cup X_2 \cup X_v$) is at most $4|X_v| = 4|I_v|$.
Thus, the number of nodes whose sets $\Delta(B)$ are adjacent to $A$ in $G-E_I$ is at most
\[
|X_1 \cup X_2| + 5 \sum_{v \in A} |I_v| = O(k^3),
\]
since $|X_1 \cup X_2| = O(k^3)$ (by Lemma~\ref{lem:X1X2}),
 $|A| = O(k)$ and $|I_v| = O(k^2)$ for each $v$.
\end{proof}

\subsection{Reducing the number of nodes in $T_{G-A-R}$}

Let $\PP$ be the set of connected components of the graph $T-X_1-X_2$.
Since $X_1$ contains all nodes in $T$ whose degree is not 2,
every connected component of $T-X_1-X_2$ is a path.
Furthermore, $|\PP| \leq |X_1|$.
Therefore, by Lemma~\ref{lem:X1X2} we conclude that $|\PP| = O(k^3)$.

For two nodes $C_1,C_2$ in the same connected component of $T$,
let $U(C_1,C_2)$ be the union of all node sets on the unique path
in $T$ from $C_1$ to $C_2$, including both endpoints.
Adjacent nodes on this path represent nested cliques, so $G[U(C_1,C_2)]$ is connected.

For a vertex $v \in A$, let $\alpha(v)$ be the number of nodes $C \in \bigcup \PP$
such that $C$ is a maximal clique in $G-A-R$ and $C$ is adjacent to $v$ in $G$.
The first reduction rule in this step aims to bound the value of $\alpha(v)$ for
every vertex $v \in A$.

\begin{rrule}\label{rule:alpha}
If there is a vertex $v \in A$ such that $\alpha(v) > 6k' + 5|\PP|$, then
set $A \gets A \setminus \{v\}$ and $E_I \gets E_I - v$,
and return the instance $(G-v,k'-1,E_M-v)$.
\end{rrule}
\begin{lemma}\label{lem:rule-alpha}
Rule~\ref{rule:alpha} is safe.
\end{lemma}
\begin{proof}
Suppose that $(G,k',E_M)$ is a yes-instance and let $S$ be a solution of this instance.
We aim to show that $v$ must be in $S$.
We will show that since $\alpha(v) > 6k' + 5|\PP|$,
we can identify a collection of $k'+1$ obstructions in $G$ such that
the intersection of the vertices of any two distinct obstructions is exactly $\{v\}$.

Let $P \in \PP$.
Let $C^P_1, \ldots C^P_{c_P}$ be the nodes in $P$ that are maximal cliques in $G-A-R$ and are adjacent to $v$
in $G$, ordered along the path $P$.
For every $i \leq \lfloor c_P/6 \rfloor$,
we select vertices $u^P_i \in C^P_{6i-5} \cap N(v)$ and $v^P_i \in C^P_{6i-1} \cap N(v)$.
Let $Q^P_i$ be a shortest path in $G[U(C^P_{6i-5},C^P_{6i-1})]$ from $u^P_i$ to $v^P_i$.
This path exists by the connectivity of this subgraph.
We have that $Q^P_i$ is an induced path in $G$ and its endpoints are adjacent to $v$.
The two endpoint cliques are separated by at least three other maximal cliques along their path in $T$.
By Lemma~\ref{lem:TG-4}, $Q^P_i$ therefore has length at least 3.
The graph $G' = G[V(Q^P_i) \cup \{v\}]$ contains a Hamiltonian cycle of length at least 5.
Additionally, every edge in $G'$ that is not on this cycle is incident to $v$.
By Lemma~\ref{lem:cycle}, the graph $G'$ contains an obstruction $H^P_i$ that contains $v$.

By Lemma~\ref{lem:TG-4}(\ref{enu:TG-CiCj}),
the sets $V(H^P_i) \setminus \{v\}$ and $V(H^P_j) \setminus \{v\}$ are disjoint for every $P \in \PP$
and $i \neq j$.
By Lemma~\ref{lem:TG-4}(\ref{enu:TG-C}),
the sets $V(H^P_i) \setminus \{v\}$ and $V(H^{P'}_j) \setminus \{v\}$ are disjoint
for every two pairs $(P,i)$ and $(P',j)$ with $P \neq P'$.
Therefore, the sets $V(H^P_i) \setminus \{v\}$ and $V(H^{P'}_j) \setminus \{v\}$ are pairwise disjoint
for every two distinct pairs $(P,i)$ and $(P',j)$.

The number of obstructions $H^P_i$ for all $P$ and $i$ is
\[ \sum_{P \in \PP} \lfloor c_P/6 \rfloor \geq \sum_{P \in \PP} \frac{1}{6}(c_P - 5)
= \frac{1}{6} (\alpha(v) - 5|\PP|) > k'.
\]
By Theorem~\ref{thm:ptolemaic-1}, $S$ hits all the obstructions in $G$,
and therefore $v \in S$.
This means that $S \setminus \{v\}$ is a solution of $(G-v,k'-1,E_M-v)$.

For the opposite direction, suppose that $S$ is a solution of $(G-v,k'-1,E_M-v)$.
Then, $S \cup \{v\}$ is a solution of $(G,k',E_M)$.
\end{proof}

\begin{lemma}\label{lem:rule-alpha-2}
Rule~\ref{rule:alpha} preserves properties~\ref{enu:aug-first}--\ref{enu:aug-last}.
\end{lemma}
\begin{proof}
Since the rule does not change the graph $G-A-R$, property~\ref{enu:aug-T} is preserved.
The remaining properties are preserved by Lemma~\ref{lem:del-v}.
\end{proof}

For a vertex $v \in A$, let $\beta(v)$ be the number of nodes $C \in \bigcup \PP$
such that $C$ is a maximal clique in $G-A-R$ and $C$ is adjacent to $R \cap N(v)$ in $G$.
The next reduction rule aims to bound the values of $\beta(v)$ for every vertex $v \in A$.

\begin{rrule}\label{rule:beta}
If there is a vertex $v \in A$ such that $\beta(v) > 7k' + 6|\PP|$, then
set $A \gets A \setminus \{v\}$ and $E_I \gets E_I - v$,
and return $(G-v,k'-1,E_M-v)$.
\end{rrule}
\begin{lemma}
Rule~\ref{rule:beta} is safe.
\end{lemma}
\begin{proof}
Suppose that $(G,k',E_M)$ is a yes-instance and let $S$ be a solution of this instance.
Let $P \in \PP$.
Let $C^P_1, \ldots C^P_{c_P}$ be the nodes in $P$ that are maximal cliques in $G-A-R$ and are adjacent to
$R \cap N(v)$ in $G$, ordered along the path $P$.
For every $1 \leq i \leq \lfloor c_P/7 \rfloor$, choose vertices $u^P_i,v^P_i \in R \cap N(v)$
adjacent to $C^P_{7i-6}$ and $C^P_{7i-2}$, respectively.
By property~\ref{enu:aug-R-N}, each of $u^P_i,v^P_i$ has a clique neighborhood in $G-A-R$.
By definition, $N(u^P_i)$ contains a vertex from $C^P_{7i-6}$
and $N(v^P_i)$ contains a vertex from $C^P_{7i-2}$.
These cliques are disjoint and non-adjacent by Lemma~\ref{lem:TG-4},
so $u^P_i \neq v^P_i$.

Let $U^P_i = U(C^P_{7i-6},C^P_{7i-2})$.
Let
\[ G^P_i = G[U^P_i \cup \{u^P_i,v^P_i\}]-\{\edge{u^P_i}{v^P_i}\}, \]
and let $Q^P_i$ be a shortest path in $G^P_i$ from $u^P_i$ to $v^P_i$.
This path exists because $G[U^P_i]$ is connected and each endpoint has a neighbor in it.
Every maximal clique on $P$ containing a neighbor of $u^P_i$ or $v^P_i$
occurs among $C^P_1,\ldots,C^P_{c_P}$, since $u^P_i,v^P_i \in R \cap N(v)$.
Since $N(u^P_i)$ contains a vertex from $C^P_{7i-6}$ and $N(u^P_i) \cap V(G-A-R)$ is a clique,
from Lemma~\ref{lem:TG-4} we obtain that
$N_{G^P_i}(u^P_i) \subseteq C^P_{7i-6} \cup C^P_{7i-5}$.
Similarly, $N_{G^P_i}(v^P_i) \subseteq C^P_{7i-3} \cup C^P_{7i-2}$.
By Lemma~\ref{lem:TG-4}, the path obtained from $Q^P_i$ by removing the endpoints $u^P_i$ and $v^P_i$
has length at least 1.
Therefore, $Q^P_i$ has length at least 3.
The vertices $u^P_i$ and $v^P_i$ are not adjacent since otherwise
$G[V(Q^P_i)]$ is a hole in $G-A$, which is a contradiction.
Since $Q^P_i$ is an induced path in $G$ whose endpoints are adjacent to $v$,
Lemma~\ref{lem:cycle} implies that $G[V(Q^P_i) \cup \{v\}]$ contains an obstruction $H^P_i$ that contains $v$.
Using the same arguments applied in the proof of Lemma~\ref{lem:rule-alpha} we obtain that
the sets $V(H^P_i) \setminus \{v,u^P_i,v^P_i\}$ and $V(H^{P'}_j) \setminus \{v,u^{P'}_j,v^{P'}_j\}$
are disjoint for every two distinct pairs $(P,i)$ and $(P',j)$.
By Lemma~\ref{lem:TG-4}, the sets $C^P_i$ and $C^P_j$ are disjoint and non-adjacent in $G$ when $|j-i| > 2$.
Additionally, the sets $C^P_i$ and $C^{P'}_j$ are disjoint and non-adjacent in $G$ if $P \neq P'$.
Since $N(u^P_i) \cap V(G-A-R)$ and $N(v^P_i) \cap V(G-A-R)$ are cliques for all $P$ and $i$,
we have that $\{u^P_i,v^P_i\} \cap \{u^{P'}_j,v^{P'}_j\} = \emptyset$
for every two distinct pairs $(P,i)$ and $(P',j)$.
Therefore, the sets $V(H^P_i) \setminus \{v\}$ and $V(H^{P'}_j) \setminus \{v\}$
are disjoint for every two distinct pairs $(P,i)$ and $(P',j)$.
The rest of the proof is similar to the proof of Lemma~\ref{lem:rule-alpha}.
\end{proof}

\begin{lemma}
Rule~\ref{rule:beta} preserves properties~\ref{enu:aug-first}--\ref{enu:aug-last}.
\end{lemma}
\begin{proof}
Since the rule does not change the graph $G-A-R$, property~\ref{enu:aug-T} is preserved.
The remaining properties are preserved by Lemma~\ref{lem:del-v}.
\end{proof}

Let $X_3$ be the set of nodes $C \in \bigcup \PP$ such that $C$ is a maximal clique in $G-A-R$
and $C$ satisfies at least one of the following conditions:
\begin{enumerate}
\item\label{X3-A}
$C$ is adjacent to $A$ in $G$.

\item\label{X3-R}
$C$ is adjacent to $R \cap N(A)$ in $G$.

\item\label{X3-EM}
$C \cap \bigcup E_M \neq \emptyset$.
\end{enumerate}
Let $\PP'$ be the set of connected components of $T - X_1 - X_2 - X_3$.

\begin{lemma}\label{lem:X3}
If Rule~\ref{rule:alpha} and Rule~\ref{rule:beta} cannot be applied, then $|X_3| = O(k^4)$.
\end{lemma}
\begin{proof}
We bound the number of nodes in $X_3$ that satisfy each of the three defining conditions separately.

The number of nodes $C \in X_3$ that satisfy (\ref{X3-A}) is at most $\sum_{v \in A} \alpha(v)$.
Since Rule~\ref{rule:alpha} cannot be applied,
$\alpha(v) \leq 6k' + 5|\PP| = O(k^3)$ for every $v \in A$.
Therefore,
$\sum_{v \in A} \alpha(v) = O(|A| \cdot k^3) = O(k^4)$.

The number of nodes $C \in X_3$ that satisfy (\ref{X3-R}) is at most $\sum_{v \in A} \beta(v)$.
Since Rule~\ref{rule:beta} cannot be applied,
$\beta(v) \leq 7k' + 6|\PP| = O(k^3)$ for every $v \in A$.
Therefore,
$\sum_{v \in A} \beta(v) = O(|A| \cdot k^3) = O(k^4)$.

Let $e \in E_M$.
By property~\ref{enu:aug-e}, at least one vertex of $e$ is in $A$.
Therefore, at most one vertex of $e$ is in $V(G-A-R)$.
If $e$ contains a vertex $v \in V(G-A-R)$, then by Lemma~\ref{lem:TG-4},
the number of nodes $C \in X_3$ that contain $v$ is at most 2.
Therefore, the number of nodes $C \in X_3$ that satisfy (\ref{X3-EM}) is at most
$2|E_M|$.
By property~\ref{enu:aug-EM}, $2|E_M| = O(k^2)$.
\end{proof}

\begin{lemma}\label{lem:Pprime}
If Rule~\ref{rule:alpha} and Rule~\ref{rule:beta} cannot be applied, then $|\PP'| = O(k^4)$.
\end{lemma}
\begin{proof}
The connected components in $\PP'$ are obtained by deleting vertices of $X_3$
from the connected components in $\PP$.
Since the connected components in $\PP$ are paths, we obtain that
$|\PP'| \leq |\PP| + |X_3|$.
Since $|\PP| = O(k^3)$ and $|X_3| = O(k^4)$ (by Lemma~\ref{lem:X3}),
we obtain that $|\PP'| = O(k^4)$.
\end{proof}

The next reduction rule aims to bound the sizes of the connected components in $T - X_1 - X_2 - X_3$.

\begin{rrule}\label{rule:path}
If $C_1,B_1,C_2,B_2,C_3,B_3,C_4,B_4,C_5$ is a path in $T-X_1-X_2-X_3$
such that $C_1,C_2,\ldots,C_5$ are maximal cliques in $G-A-R$ and $|B_3| \leq |B_2|$,
then return $(G',k',E_M)$, where $G'$ is a graph that is obtained from $G$ by deleting
the vertices in $(C_2 \cup C_3) \setminus (B_1 \cup B_3)$ and adding an edge between
every vertex in $B_1$ and every vertex in $B_3$.
\end{rrule}

Since the nodes $C_2$ and $C_3$ are not in $X_3$, they do not satisfy
condition~(\ref{X3-EM}) in the definition of $X_3$.
Namely, $(C_2 \cup C_3) \cap (\bigcup E_M) = \emptyset$.
Therefore, deleting vertices in $C_2 \cup C_3$ does not delete any endpoint of pairs in $E_M$,
so $E_M$ remains unchanged by the rule.

By property~\ref{enu:aug-e}, $A$ is a hitting set of $E_I$.
Since $C_2,C_3 \notin X_3$, they are not adjacent to $A$.
Therefore, $(C_2 \cup C_3) \cap (\bigcup E_I) = \emptyset$.
Thus, the set $E_I$ also remains unchanged.

\begin{lemma}\label{lem:Gp-A}
Let $(G',k',E_M)$ be the instance obtained by applying Rule~\ref{rule:path}.
Then, $A$ is a deletion set of $G'$.
\end{lemma}
\begin{proof}
Suppose for contradiction that $A$ is not a deletion set of $G'$.
By Theorem~\ref{thm:ptolemaic-1}, $G'-A$ contains an obstruction $H$.
Since $G'-B_1$ is an induced subgraph of $G$, we have that $H$ contains a vertex $v_1 \in B_1$
(otherwise $H$ is an obstruction in $G-A$).
Similarly, $H$ contains a vertex $v_3 \in B_3$.

Assume that $H$ is a hole.
Let $v_2$ be an arbitrary vertex from $B_2$.
Since $\edge{v_1}{v_3} \in E(G')$, the vertices $v_1$ and $v_3$ are adjacent on the hole $H$.
Therefore, the graph $G_2 = G[V(H) \cup \{v_2\}]$ contains a Hamiltonian cycle
(the original cycle of $H$ with the path $v_1,v_3$ replaced by $v_1,v_2,v_3$).
Every edge in $G_2$ that is not on this cycle is incident to $v_2$.
By Lemma~\ref{lem:cycle}, $G_2$ contains an obstruction $H'$.
Since $V(H) \subseteq V(G-A)$ and $v_2 \notin A$, we have $V(H') \subseteq V(G-A)$.
Thus, $G-A$ contains an obstruction, which is a contradiction.
Therefore, $H$ is a gem.

By Lemma~\ref{lem:TG-4}, the vertices $v_1$ and $v_3$ are nonadjacent in $G$
and their common neighbors in $G-A-R$ are exactly the vertices in $B_2$.
Furthermore, property~\ref{enu:aug-R-N} implies that
every vertex in $R$ can be adjacent to at most one of $\{v_1,v_3\}$.
Thus, the set of all common neighbors of $v_1$ and $v_3$ in $G-A$ is $B_2$.
By construction, the new edges ensure that the set of common neighbors of $v_1$ and $v_3$
in $G'-A$ is $(B_1 \cup B_3) \setminus \{v_1,v_3\}$.
Since the center of the gem $H$ must be adjacent to the other four vertices,
the center must be in $B_1 \cup B_3$.
Without loss of generality, we assume the center of $H$ is $v_1$
(we note that condition $|B_3| \leq |B_2|$ in Rule~\ref{rule:path} is not used
in the proof of the lemma, so this assumption is valid).

By Lemma~\ref{lem:TG-4}, $N_{G-A}(v_1) \subseteq R \cup C_1 \cup C_2$.
By construction,
$N_{G'-A}(v_1) \subseteq (N_{G-A}(v_1) \setminus (C_2 \setminus B_1)) \cup B_3 \subseteq R \cup C_1 \cup B_3$.
Since $v_1$ is the center of $H$,
$V(H) \setminus \{v_1\} \subseteq N_{G'-A}(v_1) \subseteq R \cup C_1 \cup B_3$.
By Theorem~\ref{thm:ptolemaic-2}, $T = \und{T_{G-A-R}}$ is acyclic.
By Lemma~\ref{lem:P-replace}, $\und{T_{G'-A-R}}$ is acyclic, and thus $G'-A-R$ is a Ptolemaic graph.
If $V(H) \cap R = \emptyset$, then $H$ is an obstruction in $G'-A-R$, which is a contradiction.
Therefore, $V(H) \cap R \neq \emptyset$.

Recall that $V(H) \subseteq V(G'-A) \subseteq V(G-A)$.
We select $v_2 \in B_2$ and $x \in C_1 \setminus B_1$.
Note that $v_2,x \in V(G-A)$.
We will show that $G[V(H) \setminus \{v_3\} \cup \{v_2,x\}]$ contains a gem,
and thus $G-A$ contains a gem.

Recall that $V(H) \subseteq R \cup C_1 \cup B_3$.
By Lemma~\ref{lem:TG-4}, $C_1 \setminus B_1$ is not adjacent to $B_3$ in $G$,
and it is also not adjacent to $B_3$ in $G'$.
We claim that $V(H) \cap R$ is not adjacent to $B_3$ in $G'$.
Let $v \in V(H) \cap R$.
We need to show that $v$ is not adjacent to $B_3$ in $G'$.
Since $v \in V(H)$ and $v_1$ is the center of $H$, $v$ is adjacent to $v_1$ in $G'$ and also in $G$.
By property~\ref{enu:aug-R-N}, $N(v) \cap V(G-A-R)$ is a clique in $G$.
Therefore, $v$ cannot be adjacent to any non-neighbor of $v_1$ in $G-A-R$.
By Lemma~\ref{lem:TG-4}, $v_1$ is not adjacent to $B_3$ in $G$.
Therefore, $v$ is not adjacent to $B_3$ in $G$.
Since the construction of $G'$ only adds edges between $B_1$ and $B_3$,
$v$ is not adjacent to $B_3$ in $G'$. Thus, the claim holds.

Since $H$ is a gem, the removal of the center $v_1$ leaves an induced path of length~3.
We use this to constrain the remaining vertices of $H$.
Since $H$ contains a vertex in $R$ and a vertex in $B_3$ (the vertex $v_3$),
from the previous paragraph we have that $H$ contains a vertex $u_1 \in B_1 \setminus \{v_1\}$.
If $H$ contains a vertex $w \in B_3 \setminus \{v_3\}$, then $\{v_3,u_1,w\}$ would form a clique in $H$.
This contradicts the fact that $H-v_1$ is an induced path.
Therefore, $V(H) \cap B_3 = \{v_3\}$.
Similarly, $V(H) \cap B_1 = \{v_1,u_1\}$ otherwise $H-v_1$ contains a clique of size 3.
We showed above that the vertices of $H$ in $R \cup (C_1 \setminus B_1)$ are not adjacent to $B_3$ in $G'$.
Thus, $v_3$ is an endpoint of $H-v_1$ and its neighbor on this path is $u_1$.
The induced path $H-v_1$ must therefore be one of two forms below (see Figure~\ref{fig:rule-path}):
\begin{enumerate}
\item
A path $r,u,u_1,v_3$ where $r \in R$ and $u \in C_1 \setminus B_1$.
\item
A path $r,r',u_1,v_3$ where $r,r' \in R$.
\end{enumerate}

\begin{figure}
\centering
\subfigure[]{\includegraphics{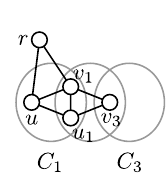}}%
\qquad%
\subfigure[]{\includegraphics{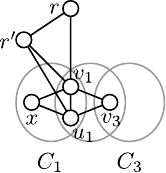}}%
\caption{The two cases of the gem $H$.
Figure~(a) shows the case $H-v_1$ is a path $r,u,u_1,v_3$ where $r \in R$ and $u \in C_1 \setminus B_1$,
and figure~(b) shows the case $H-v_1$ is a path $r,r',u_1,v_3$ where $r,r' \in R$.\label{fig:rule-path}}
\end{figure}

\textbf{Case 1:} $H-v_1$ is a path $r,u,u_1,v_3$.\\
Assume the induced path $H-v_1$ is $r,u,u_1,v_3$, where $r \in R$ and $u \in C_1 \setminus B_1$.
By the construction of $G'$, $G[\{v_1,r,u,u_1\}] = G'[\{v_1,r,u,u_1\}]$.

We now consider the edges between $v_2$ and $\{v_1,r,u,u_1\}$ in $G$.
By Lemma~\ref{lem:TG-4}, $v_2$ is adjacent to both $v_1$ and $u_1$ in $G$,
and not adjacent to $u$ (since $u \in C_1 \setminus B_1$).
By property~\ref{enu:aug-R-N}, $N(r) \cap V(G-A-R)$ is clique in $G$.
Since $\edge{r}{u} \in E(G)$, $r$ cannot be adjacent to any non-neighbor of $u$ in $G-A-R$.
Therefore, $v_2$ is not adjacent to $r$ in $G$.
Thus, $G[\{v_1,r,u,u_1,v_2\}]$ is a gem in $G-A$, with center $v_1$
and induced path $r,u,u_1,v_2$, which is a contradiction.

\textbf{Case 2:} $H-v_1$ is a path $r,r',u_1,v_3$.\\
Assume the induced path $H-v_1$ is $r,r',u_1,v_3$ where $r,r' \in R$.
Our objective is to show that the subgraph $G[V(H) \setminus \{v_3\} \cup \{v_2,x\} ]$ contains a gem with a center $v_1$.
The vertex $v_1$ is adjacent to $r$, $r'$, $u_1$, $v_2$, and $x$ in $G$.
Let $Y = \{r,r',u_1,v_2,x\}$.
Thus, it suffices to show that $G[Y]$ contains an induced path of length 3.

By construction, $G[\{r,r',u_1\}] = G'[\{r,r',u_1\}]$ and thus
$\edge{r}{r'}, \edge{r'}{u_1} \in E(G)$ and $\edge{r}{u_1} \notin E(G)$.
Since $x \in C_1 \setminus B_1$ and $v_2 \in B_2$, we have $\edge{x}{v_2} \notin E(G)$.
Since $N(r) \cap V(G-A-R)$ and $N(r') \cap V(G-A-R)$ are cliques in $G$ and $\edge{x}{v_2} \notin E(G)$,
we have $|N(r) \cap \{x,v_2\}| \leq 1$ and $|N(r') \cap \{x,v_2\}| \leq 1$.

Consider the graph $G[\{r,x,u_1,v_2\}]$.
If $\edge{r}{x} \in E(G)$, then by the inequality $|N(r) \cap \{x,v_2\}| \leq 1$,
we must have $\edge{r}{v_2} \notin E(G)$.
Therefore, $r,x,u_1,v_2$ is an induced path and we are done.
Symmetrically, if $\edge{r}{v_2} \in E(G)$, then $\edge{r}{x} \notin E(G)$
and $r,v_2,u_1,x$ is an induced path.
We now assume $\edge{r}{x},\edge{r}{v_2} \notin E(G)$.

Consider the graph $G[\{r,r',u_1,x\}]$.
We know that $r,r',u_1$ is an induced path and $\edge{r}{x} \notin E(G)$.
If $\edge{r'}{x} \notin E(G)$, then $r,r',u_1,x$ is an induced path.
We now assume $\edge{r'}{x} \in E(G)$.

Finally, consider the graph $G[\{r,r',u_1,v_2\}]$.
Since $|N(r') \cap \{x,v_2\}| \leq 1$ and $\edge{r'}{x} \in E(G)$,
we have $\edge{r'}{v_2} \notin E(G)$.
Therefore, $r,r',u_1,v_2$ is an induced path.

Since we found an induced path in every possible case, the graph $G[Y \cup \{v_1\}]$
contains a gem with center $v_1$, and this gem is contained in $G-A$.
The existence of an obstruction in $G-A$ is a contradiction to property~\ref{enu:aug-A}.
Therefore, $A$ is a deletion set of $G'$.
\end{proof}

\begin{lemma}
Rule~\ref{rule:path} is safe.
\end{lemma}
\begin{proof}
Suppose that $(G,k',E_M)$ is a yes-instance, and let $S$ be a solution of this instance.
Let $\Vd = (C_2 \cup C_3) \setminus (B_1 \cup B_3)$ be the set of vertices deleted by the rule.
Define a set $S'$ as follows:
\[
S' = \begin{cases}
S \setminus \Vd          & \text{if } |S \cap B_2| < |B_3|\\
(S \setminus \Vd) \cup B_3 & \text{if } |S \cap B_2| \geq |B_3|.
\end{cases}
\]
In both cases, $S'\subseteq V(G')$.
We will show that $S'$ is a solution of $(G',k',E_M)$.

In the first case of the definition of $S'$, $|S'| \leq |S|$.
In the second case we have
$|S'| \leq |S|-|S \cap \Vd|+|B_3| \leq |S|-|S \cap B_2|+|B_3| \leq |S|$.
Thus, we always have $|S'| \leq |S| \leq k'$.

Since $C_2,C_3 \notin X_3$, $(C_2 \cup C_3) \cap \bigcup E_M = \emptyset$.
As $S \setminus S' \subseteq \Vd \subseteq C_2 \cup C_3$,
$(S \setminus S') \cap \bigcup E_M = \emptyset$.
Since $S$ is a hitting set of $E_M$, we have that $S'$ is a hitting set of $E_M$.

It remains to show that $S'$ is a deletion set of $G'$.
Suppose for contradiction that $S'$ is not a deletion set of $G'$.
By Theorem~\ref{thm:ptolemaic-1}, $G'-S'$ contains an obstruction $H$.
If $V(H) \cap B_1 = \emptyset$ or $V(H) \cap B_3 = \emptyset$, then
$H$ is an obstruction in $G-S$, contradicting that $S$ is a solution.
Therefore, $H$ contains a vertex $v_1 \in B_1$ and a vertex $v_3 \in B_3$.

By Lemma~\ref{lem:Gp-A}, $A$ is a deletion set of $G'$.
By Lemma~\ref{lem:P-replace} (on the graph $G-A-R$), 
$C_1,B_1,B_1 \cup B_3,B_3,C_4$ is a path in $\und{T_{G'-A-R}}$
that satisfies the conditions of Lemma~\ref{lem:path-2}
(note that $B_1 \cup B_3$ is not adjacent to $A \cup (R \cap N_{G'}(A)) = A \cup (R \cap N(A))$ in $G'$
since $C_2$ and $C_3$ are not adjacent to $A \cup (R \cap N(A))$ in $G$).
By Lemma~\ref{lem:path-2}, $H$ is a hole.
Since $v_3 \in V(H)$, we have $v_3 \notin S'$.
If the second case in the definition of $S'$ held,
then $v_3 \in B_3 \subseteq S'$, which is a contradiction.
Therefore, we must be in the first case, so $|S \cap B_2| < |B_3| \leq |B_2|$.
Since $|S \cap B_2| < |B_2|$, there is a vertex $v_2 \in B_2 \setminus S$.
By Lemma~\ref{lem:cycle}, the graph $G[V(H) \cup \{v_2\}]$ contains an obstruction $H'$.
Since $H$ is an obstruction in $G'-S'$, we have $V(H) \cap S' = \emptyset$
and $V(H) \cap \Vd = \emptyset$ (as $G'$ does not contain vertices from $\Vd$).
Therefore, $V(H) \cap S \subseteq V(H) \cap (S' \cup \Vd) = \emptyset$.
Since $v_2 \notin S$ and $V(H') \subseteq V(H) \cup \{v_2\}$,
we have $V(H') \cap S = \emptyset$.
Thus, $H'$ is also an obstruction in $G-S$, which is a contradiction.
Therefore, $S'$ is a deletion set of $G'$, and $(G',k',E_M)$ is a yes-instance.

For the opposite direction, suppose that $(G',k',E_M)$ is a yes-instance,
and let $S$ be a solution of this instance.
We will show that $S$ is also a solution of $(G,k',E_M)$.
Since $|S| \leq k'$ and $S$ is a hitting set of $E_M$, it remains to show that
$S$ is a deletion set of $G$.
Suppose for contradiction that $S$ is not a deletion set of $G$, and let
$H$ be an obstruction in $G-S$.

Since $G-(C_2 \cup C_3)$ is an induced subgraph of $G'$,
we have that $H$ contains a vertex from $C_2 \cup C_3$.
Suppose that $H$ contains a vertex from $C_2$.
By Lemma~\ref{lem:path-2} on the path $C_1,B_1,C_2,B_2,C_3$ in $T$,
$H$ is a hole.
Since $S$ is a hitting set of $E_M$, by properties~\ref{enu:aug-EI} and~\ref{enu:aug-R-EI},
$H$ does not contain a vertex from $R$,
so $H$ is a hole in the graph $G_1 = G-R$.
By Lemma~\ref{lem:path-1} on the graph $G_1$ and the path $C_1,B_1,C_2,B_2,C_3$ in $\und{T_{G_1-A}} = T$,
$|V(H) \cap B_1| = 1$, $|V(H) \cap B_2| = 1$, and $V(H) \cap (C_2 \setminus (B_1 \cup B_2)) = \emptyset$.
By Lemma~\ref{lem:path-1} on the graph $G_1$ and the path $C_2,B_2,C_3,B_3,C_4$ in $\und{T_{G_1-A}} = T$,
$|V(H) \cap B_3| = 1$ and $V(H) \cap (C_3 \setminus (B_2 \cup B_3)) = \emptyset$.
Thus, the construction of $G'$ ensures that $H' = G'[V(H) \setminus B_2]$ is a chordless cycle in $G'$.
By Lemma~\ref{lem:path-1} we also have $|V(H) \cap B_4| = 1$.
Since $H$ must contain a vertex from $A$, it follows that $|V(H)| \geq 5$.
Therefore, the length of the cycle $H'$ is $|V(H)|-1 \geq 4$.
Since $V(H) \cap S = \emptyset$, $H'$ is a hole in $G'-S$, which is a contradiction.

If $H$ contains a vertex from $C_3$, we obtain a contradiction using a similar proof.
The difference is that we first apply Lemma~\ref{lem:path-1} on the path $C_2,B_2,C_3,B_3,C_4$
and then on the path $C_1,B_1,C_2,B_2,C_3$.

Therefore, $S$ is a deletion set of $G$, and $(G,k',E_M)$ is a yes-instance.
\end{proof}

\begin{lemma}
Rule~\ref{rule:path} preserves properties~\ref{enu:aug-first}--\ref{enu:aug-last}.
\end{lemma}
\begin{proof}
The rule does not modify $k'$, $A$, $E_I$, $E_M$, and $R$.
Therefore, it preserves properties~\ref{enu:aug-kp}, \ref{enu:aug-e}, \ref{enu:aug-EM},
and~\ref{enu:aug-R-size}.
By Lemma~\ref{lem:Gp-A}, the rule preserves property~\ref{enu:aug-A}.
We now show it preserves the rest of the properties.

(\ref{enu:aug-EI}) Let $K$ be a hole in $G'$ that contains an edge $e_I \in E_I$.
If $V(K) \cap B_1 = \emptyset$ or $V(K) \cap B_3 = \emptyset$, then $K$
does not use the new edges introduced by the rule.
Therefore $K$ is also a hole in $G$.
By property~\ref{enu:aug-EI} for $G$,  there is a pair $e \in E_M$ such that
$e \subseteq V(K)$.

Suppose $V(K) \cap B_1 \neq \emptyset$ and $V(K) \cap B_3 \neq \emptyset$.
Since $B_1 \cup B_3$ is a clique in $G'$ (due to the new edges),
$|V(K) \cap (B_1 \cup B_3)| \leq 2$, and therefore
$|V(K) \cap B_1| = 1$ and $|V(K) \cap B_3| = 1$.
Let $v_1$ and $v_3$ be the unique vertices in $V(K) \cap B_1$ and $V(K) \cap B_3$, respectively.

Our goal is to find a hole $K_3$ in $G$ that contains $e_I$.
Arbitrarily pick a vertex $v_2 \in B_2$.
Let $K_2 = G[V(K) \cup \{v_2\}]$.
The graph $K_2$ contains a Hamiltonian cycle whose edges are
$E(K) \setminus \{\edge{v_1}{v_3}\} \cup\{\edge{v_1}{v_2},\edge{v_2}{v_3}\}$.
Every edge of $K_2$ that does not belong to this cycle is incident to $v_2$.
By property~\ref{enu:aug-e}, $e_I \cap A \neq \emptyset$, so let $w \in e_I \cap A$.
Since $v_2 \in C_2$ and $C_2 \notin X_3$, we have that $v_2$ is not adjacent to $A$ in $G$,
so $\edge{v_2}{w} \notin E(G)$.
By Lemma~\ref{lem:cycle-2}, there is a hole $K_3$ in $G$ such that
$e_I \cup \{v_2\} \subseteq V(K_3) \subseteq V(K_2) = V(K) \cup \{v_2\}$.
Since $K_3$ is a hole in $G$ containing $e_I \in E_I$,
property~\ref{enu:aug-EI} for $G$ guarantees a pair $e \in E_M$ such that $e \subseteq V(K_3)$.
Since $C_2 \notin X_3$, we have $C_2 \cap \bigcup E_M = \emptyset$.
Because $v_2 \in C_2$ and $e \in E_M$, we have that $v_2 \notin e$.
As $e \subseteq V(K_3) \subseteq V(K) \cup \{v_2\}$ and $v_2 \notin e$,
we conclude $e \subseteq V(K)$.
Therefore, property~\ref{enu:aug-EI} is preserved.

(\ref{enu:aug-R-EI})
Let $K$ be a hole in $G'$ that contains a vertex from $R$.
If $V(K) \cap B_1 = \emptyset$ or $V(K) \cap B_3 = \emptyset$, then $K$ is a hole in $G$
that contains a vertex from $R$.
By property~\ref{enu:aug-R-EI}, $K$ contains an edge from $E_I$.

Suppose $V(K) \cap B_1 \neq \emptyset$ and $V(K) \cap B_3 \neq \emptyset$.
We define vertices $v_1,v_2,v_3$ and a graph $K_2 = G[V(K) \cup \{v_2\}]$ as in the proof for
property~\ref{enu:aug-EI}.
Since $A$ is a deletion set of $G'$, $K$ contains a vertex $w \in A$.
As $v_2$ is not adjacent to $A$ (because $v_2 \in C_2$ and $C_2 \notin X_3$),
$\edge{v_2}{w} \notin E(G)$.
By Lemma~\ref{lem:cycle-2},  there is a hole $K_3$ in $G$ such that
$\{v_2\} \subseteq V(K_3) \subseteq V(K_2)$.

We claim that $K_3$ must contain a vertex from $R$.
If $K_3 = K_2$, then $V(K) \subseteq V(K_3)$, so $K_3$ contains the original vertex from $R$.
Otherwise, $K_3$ utilizes an edge $\edge{v_2}{x}$, where $x \in V(K) \setminus \{v_1,v_3\}$.
Since $v_2$ is not adjacent to $A$, by Lemma~\ref{lem:TG-4},
$N[v_2] \subseteq N_{G-A-R}[v_2] \cup R = C_2 \cup C_3 \cup R$.
The hole $K$ only intersects $C_2 \cup C_3$ at $v_1$ and $v_3$ (using the same argument as
in the proof for property~\ref{enu:aug-EI}).
Therefore, $x \in R$.
In both cases, $K_3$ contains a vertex from $R$.

Since $K_3$ is a hole in $G$ and contains a vertex from $R$,
property~\ref{enu:aug-R-EI} for $G$ implies that $K_3$ contains an edge $e \in E_I$.
By property~\ref{enu:aug-e}, $e \cap A \neq \emptyset$.
Since $v_2$ is not adjacent to $A$, $v_2 \notin e$.
Because $e \subseteq V(K_3) \subseteq V(K) \cup \{v_2\}$,
we conclude that $e \subseteq V(K)$.
Therefore, property~\ref{enu:aug-R-EI} is preserved.

(\ref{enu:aug-R-N})
Let $v \in R$.
By property~\ref{enu:aug-R-N} for $G$, $N(v) \cap V(G-A-R)$ is a clique in the original graph $G$.
By construction, $V(G')$ differs from $V(G)$ by the deleted vertices $(C_2 \cup C_3) \setminus (B_1 \cup B_3)$.
Furthermore, the graph $G'$ differs from $G[V(G')]$ by the new edges between $B_1$ and $B_3$.
Since $v \in R$, $v$ is not an endpoint of a new edge.
Therefore, $N_{G'}(v) \cap V(G'-A-R) \subseteq N(v) \cap V(G-A-R)$.
Since a subset of a clique is also a clique, $N_{G'}(v) \cap V(G'-A-R)$ is a clique in $G$.
Since the graph $G'$ differs from $G[V(G')]$ by the new edges,
we obtain that $N_{G'}(v) \cap V(G'-A-R)$ is a clique in $G'$.
Therefore, property~\ref{enu:aug-R-N} is preserved.

(\ref{enu:aug-dI})
Let $v \in A$.
Let $I'$ be a maximum independent set in the graph $G'[N_{G'-E_I}(v)]$.
Since $V(G') \subset V(G)$, $I' \subseteq V(G)$.
Since the reduction rule only adds edges between $B_1$ and $B_3$ and
$v \in A$ is not an endpoint of a new edge,
$I' \subseteq N_{G'-E_I}(v) \subseteq N_{G-E_I}(v)$.
Every edge of $G$ whose endpoints are retained remains in $G'$.
Therefore, $I'$ is an independent set in $G[N_{G-E_I}(v)]$.
Thus, $d^I_{G',E_I}(v) = |I'| \leq d^I_{G,E_I}(v)$.
By property~\ref{enu:aug-dI} for $G$, $d^I_{G,E_I}(v) = O(k^2)$.
Therefore, $d^I_{G',E_I}(v) = O(k^2)$ and property~\ref{enu:aug-dI} is preserved.

(\ref{enu:aug-T})
By Lemma~\ref{lem:P-replace}, the number of leaves in $T_{G'-A-R}$ is equal to the
number of leaves in $T_{G-A-R}$, and thus property~\ref{enu:aug-T} is maintained.

\end{proof}

\begin{lemma}\label{lem:T}
If Rules~\ref{rule:alpha}, \ref{rule:beta} and~\ref{rule:path} cannot be applied, then $|V(T)| = O(k^4)$.
\end{lemma}
\begin{proof}
By definition, $V(T) = X_1 \cup X_2 \cup X_3 \cup \bigcup \PP'$.
By Lemma~\ref{lem:X1X2} and Lemma~\ref{lem:X3},
$|X_1 \cup X_2 \cup X_3| = O(k^4)$.
If a path $P \in \PP'$ has at least ten nodes, it contains a subpath
$C_1,B_1,C_2,B_2,C_3,B_3,C_4,B_4,C_5$ whose $C_i$ are maximal cliques.
Either $|B_3| \leq |B_2|$, or this inequality holds after reversing the subpath,
so Rule~\ref{rule:path} applies.
Since that rule cannot be applied, $|P| \leq 9$ for every $P \in \PP'$.
By Lemma~\ref{lem:Pprime}, $|\PP'| = O(k^4)$.
Therefore,
$|\bigcup \PP'| \leq 9 \cdot |\PP'| = O(k^4)$.
\end{proof}

\subsection{Reducing the number of vertices in  $G-A-R$}

Let $\D$ be the set of all sets $X \subseteq V(G)$ such that $|X| \leq 5$ and $G[X]$ is an obstruction.
The kernelization algorithm applies the algorithm of Theorem~\ref{thm:hitting-set} on the instance
$(\D,k')$, to obtain the set $\D'$.
Let $D = \bigcup \D'$.

For the last reduction rule we use a marking procedure that identifies a set of important
vertices in $G$.
When we say that the procedure marks $p$ vertices from a set $X$,
this means it marks exactly $\min(p,|X|)$ vertices.
The marking procedure is as follows:

\begin{enumerate}
\item\label{mark-A}
Mark every vertex in $A \cup R$.

\item\label{mark-D}
Mark every vertex in $D$.

\item\label{mark-EM}
Mark every vertex in $\bigcup E_M$.

\item\label{mark-AA}
For every $s_1,s_2 \in A$, mark $k+1$ vertices in
$(N_G(s_1) \cap N_G(s_2)) \setminus A$.

\item\label{mark-B}
For every $B \in V(T)$, mark $k+1$ vertices in $\Delta(B)$.

\item\label{mark-AB}
For every $B \in V(T)$ and $s \in A$, mark $k+2$ vertices in $N_{G-E_I}(s) \cap \Delta(B)$.
\end{enumerate}

We note that the marking procedure is applied only once,
and afterward, the kernelization algorithm stops applying the previous reduction rules
and applies Rule~\ref{rule:unmarked} below once, if possible.

\begin{lemma}\label{lem:mark}
The marking procedure marks $O(k^5)$ vertices.
\end{lemma}
\begin{proof}
We have
$|A| = O(k)$,
$|R| = O(k^5)$ (by property~\ref{enu:aug-R-size}),
$|D| \leq 5|\D'| = O((k')^5) = O(k^5)$ (by Theorem~\ref{thm:hitting-set}),
$|\bigcup E_M| = O(k^2)$ (by property~\ref{enu:aug-EM}), and
$|V(T)|=O(k^4)$ (by Lemma~\ref{lem:T}).
Therefore, step~\ref{mark-A} marks $O(k^5)$ vertices,
step~\ref{mark-D} marks $O(k^5)$ vertices,
step~\ref{mark-EM} marks $O(k^2)$ vertices,
step~\ref{mark-AA} marks $O(|A|^2 k) = O(k^3)$ vertices, and
step~\ref{mark-B} marks $O(|V(T)| \cdot k) = O(k^5)$ vertices.
Additionally, by Lemma~\ref{lem:T-A}, only $O(k^3)$ nodes $B$ have a set $\Delta(B)$
with a neighbor in $A$ in $G-E_I$.
Thus, step~\ref{mark-AB} marks $O(k^3 \cdot |A| \cdot k) = O(k^5)$ vertices.
\end{proof}

\begin{rrule}\label{rule:unmarked}
Let $X$ be the set of all unmarked vertices. If $X \neq \emptyset$, then set $E_I \gets \{e \in E_I : e \cap X = \emptyset\}$ and return $(G-X,k',E_M)$.
\end{rrule}

Note that Rule~\ref{rule:unmarked} does not modify $E_M$ since $X \cap \bigcup E_M = \emptyset$
(due to step~\ref{mark-EM} in the marking procedure).

\begin{lemma}
Rule~\ref{rule:unmarked} is safe.
\end{lemma}
\begin{proof}
Let $X = \{x_1,\ldots,x_t\}$ be the set of all unmarked vertices, and let $X_0 = \emptyset$ and $X_j = \{x_1,\ldots,x_j\}$ for $1 \leq j \leq t$.
Suppose for contradiction that $(G-X,k',E_M)$ is not equivalent to $(G,k',E_M)$.
If $S$ is a solution of $(G,k',E_M)$, then $S \setminus X$ is a solution of $(G-X,k',E_M)$,
since the Ptolemaic property is hereditary and $X \cap \bigcup E_M = \emptyset$.
Thus, $(G-X,k',E_M)$ is a yes-instance while $(G,k',E_M)$ is a no-instance.
Let $i$ be the smallest index such that $(G-X_i,k',E_M)$ is a yes-instance, and set $G_2 = G-X_{i-1}$ and $x = x_i$.
By minimality of $i$, $(G_2,k',E_M)$ is a no-instance, whereas $(G_2-x,k',E_M)$ is a yes-instance.
Let $S$ be a solution of $(G_2-x,k',E_M)$. Since $|S| \leq k'$ and $S$ is a hitting set of $E_M$, $G_2-S$ is not a Ptolemaic graph.
By Theorem~\ref{thm:ptolemaic-1}, $G_2-S$ contains an obstruction $H$, which is also an induced subgraph of $G$.
Every marked vertex belongs to $V(G_2-x)$. In particular, $D \subseteq V(G_2-x)$.
This means that every set in $\D'$ induces an obstruction in $G_2-x$.
Therefore, $S$ is a hitting set of $\D'$.
By Theorem~\ref{thm:hitting-set}, $S$ is also a hitting set of $\D$.
Thus, $V(H) \notin \D$, which means that $H$ is a hole with at least 6 vertices.
Since $S$ is a deletion set of $G_2-x$, we have $x \in V(H)$.

Due to step~\ref{mark-A} of the marking procedure,
the vertices in $A \cup R$ are marked.
Since $x$ is unmarked, $x \notin A \cup R$.
Thus, $x \in V(G-A-R)$.
Let $B$ be the node of $T$ such that $x \in \Delta(B)$
($B$ exists and is unique by Lemma~\ref{lem:delta-C-0}).
Let $u,v$ be the two neighbors of $x$ in the hole $H$.
Since $H$ is a hole and $\Delta(B)$ is a clique,
$(V(H) \setminus \{x,u,v\}) \cap \Delta(B) = \emptyset$.

Suppose $H$ contains an edge from $E_I$.
By property~\ref{enu:aug-EI}, there is a pair $e \in E_M$ such that $e \subseteq V(H)$.
Since $S$ is a hitting set of $E_M$, we obtain that $V(H) \cap S \neq \emptyset$,
which contradicts the assumption that $G_2-S$ contains the graph $H$.
Therefore, $H$ does not contain an edge from $E_I$.
If $H$ contained a vertex from $R$, by property~\ref{enu:aug-R-EI},
$H$ would contain an edge from $E_I$, which is a contradiction.
Therefore, $H$ does not contain any vertex from $R$.
In particular, $u,v \notin R$.
We consider three cases.

\textbf{Case 1:} $u,v \in A$.\\
Since $x$ is unmarked, step~\ref{mark-AA} marks $k+1$ vertices in $(N_G(u) \cap N_G(v)) \setminus A$.
Since $|S| \leq k' \leq k$ and $k+1$ vertices are marked,
at least one of these marked vertices is not in $S$.
Let $x'$ be such a vertex.

Since $H$ is a hole with at least 6 vertices, the vertex $x$ is the unique
common neighbor of $u$ and $v$ in $H$.
Since $x'$ is adjacent to both $u$ and $v$ and $x' \neq x$,
we have $x' \notin V(H)$.

We define the induced subgraph $H' = G[V(H) \setminus \{x\} \cup \{x'\}]$.
Since $x'$ is adjacent to both $u$ and $v$,
$H'$ contains a Hamiltonian cycle
(the original cycle of $H$ with the path $u,x,v$ replaced by $u,x',v$).
Additionally, every edge in $H'$ that is not on this cycle is incident to $x'$.
By Lemma~\ref{lem:cycle}, $H'$ contains an obstruction.
The graph $H'$ is an induced subgraph of $(G_2-x)-S$ (since $x'$ replaces $x$ and $x' \notin S$).
This means that $(G_2-x)-S$ contains an obstruction, which contradicts the assumption
that $S$ is a solution of $(G_2-x,k',E_M)$.

\textbf{Case 2:} $u,v \notin A$.\\
Since $u,v \notin R$, we have $u,v \in V(G-A-R)$.
We claim that $V(H) \cap \Delta(B) = \{x\}$.
Suppose for contradiction that $|V(H) \cap \Delta(B)| \geq 2$.
Since we previously established that $(V(H) \setminus \{x,u,v\}) \cap \Delta(B) = \emptyset$,
either $u \in \Delta(B)$ or $v \in \Delta(B)$.
Suppose without loss of generality that $u \in \Delta(B)$.
Since $\Delta(B)$ is a clique, $v \notin \Delta(B)$
otherwise $\edge{u}{v}$ is a chord in $H$, which contradicts the fact that $H$ is a hole.
By Lemma~\ref{lem:delta-C-2}, $\Delta(B)$ is a module in $G-A-R$.
Since $x,u \in \Delta(B)$ and $x$ is adjacent to $v$,
the module property implies that $u$ is also adjacent to $v$.
Thus $\edge{u}{v}$ is a chord in $H$, which is a contradiction.
Therefore, $V(H) \cap \Delta(B) = \{x\}$.

Since $x$ is unmarked, step~\ref{mark-B} marks $k+1$ vertices in $\Delta(B)$.
Let $x'$ be one of these marked vertices such that $x' \notin S$
($x'$ must exist since $|S| \leq k$).
We know $x' \notin V(H)$ since $x' \neq x$ and $V(H) \cap \Delta(B) = \{x\}$.
Since $x,x'$ are vertices in the module $\Delta(B)$,
$u,v$ are vertices in $G-A-R$ that are outside $\Delta(B)$, and $x$ is adjacent to both $u$ and $v$,
the module property implies that $x'$ is adjacent to both $u$ and $v$.
As in Case 1, the induced subgraph $H' = G[V(H) \setminus \{x\} \cup \{x'\}]$ contains
a Hamiltonian cycle with $u,x',v$ replacing $u,x,v$.
By Lemma~\ref{lem:cycle}, $H'$ contains an obstruction,
and therefore $(G_2-x)-S$ contains an obstruction, which is a contradiction.

\textbf{Case 3:} $|A \cap \{u,v\}| = 1$.\\
Suppose without loss of generality that $u \in A$ and $v \notin A$.
We have $x \in N_{G-E_I}(u)$ because $H$ contains no edges from $E_I$.
Since $x$ is unmarked, step~\ref{mark-AB} marks $k+2$ vertices in $N_{G-E_I}(u) \cap \Delta(B)$.
Since $|S \cup \{v\}| \leq k+1$, we can find a vertex $x'$ from
these marked vertices such that $x' \notin S \cup \{v\}$.
Since $(V(H) \setminus \{x,u,v\}) \cap \Delta(B) = \emptyset$,
$u \notin \Delta(B)$, and $x' \notin \{x,v\}$,
we have that $x' \notin V(H)$.
Since $x,x' \in \Delta(B)$, $\Delta(B)$ is a clique and a module in $G-A-R$,
and $x$ is adjacent to $v$, we have that $x'$ is adjacent to $v$.
As in the previous cases, the induced subgraph $H' = G[V(H) \setminus \{x\} \cup \{x'\}]$ 
contains an obstruction, and therefore $(G_2-x)-S$ contains an obstruction, which is a contradiction.

In all three possible cases for the neighbors of $x$ we obtained a contradiction.
Therefore, $(G-X,k',E_M)$ and $(G,k',E_M)$ are equivalent.
\end{proof}

The structural properties of Lemma~\ref{lem:augmented} are no longer needed
after Rule~\ref{rule:unmarked} is applied.
The rule leaves $k'$ and $E_M$ unchanged, so the bounds $k' \leq k$
and $|E_M| = O(k^2)$ remain valid for the conversion to an instance of
\textsc{Ptolemaic Vertex Deletion}.

We obtain the following theorem.
\begin{theorem}
There is a kernel for \textsc{Ptolemaic Vertex Deletion} with $O(k^5)$ vertices.
\end{theorem}
\begin{proof}
Given an instance $(G_0,k)$ of \textsc{Ptolemaic Vertex Deletion},
the kernelization algorithm first handles the trivial parameter values described in the overview
and computes a deletion set $A$ using Theorem~\ref{thm:approximation}.
If $|A| > 68k$, it returns a fixed no-instance.
Otherwise, it applies Lemma~\ref{lem:augmented}
to generate an equivalent instance
of \textsc{Augmented Ptolemaic Vertex Deletion}.
If the preprocessing concludes that the instance is a no-instance, the algorithm returns a fixed no-instance.
The algorithm then exhaustively applies Rules~\ref{rule:alpha}--\ref{rule:path},
recomputing the auxiliary inter-clique graph and its node sets as needed.
Whenever $k' \leq 0$, it returns a fixed instance as described in the overview.
It next computes $\D'$ and performs the marking procedure once,
and then applies Rule~\ref{rule:unmarked} once, if possible.
At this point, the current graph has $O(k^5)$ vertices
(by Lemma~\ref{lem:mark}).
Finally, the algorithm reduces the current instance of \textsc{Augmented Ptolemaic Vertex Deletion}
back to the original \textsc{Ptolemaic Vertex Deletion} problem.
As described in Step~5, this adds $k+2$ paths of length three for each pair in $E_M$,
using $2(k+2)|E_M|=O(k|E_M|)$ new vertices.
Since $|E_M| = O(k^2)$, this step adds $O(k^3)$ vertices.
Therefore, the resulting instance has $O(k^5+k^3) = O(k^5)$ vertices.
Each reduction rule strictly decreases the number of vertices, so there are at most $|V(G_0)|$ applications.
All preprocessing, marking, and reduction steps run in polynomial time,
and $k<|V(G_0)|$ for the instances considered by the algorithm.
Thus, the entire kernelization algorithm run in polynomial time in $|V(G_0)|$.
\end{proof}

\bibliographystyle{plainnat}
\bibliography{ptolemaic,dekel}

\end{document}